\documentclass[12pt, draftclsnofoot, onecolumn]{IEEEtran}
\usepackage{graphicx}
\usepackage{amsmath}
\usepackage{algorithm}
\usepackage[noend]{algorithmic}
\usepackage{amsfonts}
\usepackage{amssymb}
\usepackage{bm}
\usepackage{cite}
\usepackage{color}
\usepackage{multirow}
\usepackage{tabularx}
\newtheorem{lem}{Lemma}
\newtheorem{prob}{Problem}
\newcolumntype{L}[1]{>{\raggedright\arraybackslash}p{#1}}
\newcolumntype{C}[1]{>{\centering\arraybackslash}p{#1}}
\newcolumntype{R}[1]{>{\raggedleft\arraybackslash}p{#1}}

\usepackage{times, epsfig}
\usepackage{subfig}
\newlength{\figwidth}
\makeatletter
\renewcommand{\maketag@@@}[1]{\hbox{\m@th\normalsize\normalfont#1}}%
\makeatother

\begin{document}


\title{Energy-Efficient Task Offloading and Resource Allocation for Multiple Access Mobile Edge Computing}

\author{
Bizheng Liang, Rongfei Fan, {\it Member, IEEE}, Han Hu, {\it Member, IEEE},
\thanks
{
B. Liang, R. Fan, and H. Hu are with the School of Information and Electronics, Beijing Institute of Technology, Beijing 100081, P. R. China. (\{liangbizheng, fanrongfei, hhu\}@bit.edu.cn,).
}

}
\maketitle


\begin{abstract}
In this paper, the problem of joint radio and computation resource management over multi-channel is investigated for multi-user partial offloading mobile edge computing (MEC) system.
The target is to minimize the weighted sum of energy consumption
by jointly optimizing transmission time, local and edge computation capacity allocation, bandwidth allocation and data partition.
An optimization problem is formulated, which is nonconvex and can not be solved directly.
Then, we transform the origin optimization problem into an equivalent convex optimization problem.
%
For general case of multi-user multi-channel, we decouple the convex optimization problem into subproblems and an optimal resource management strategy is obtained by adopting block coordinate descent (BCD) method.
To gain further insight, we investigate the optimal resource management strategy for two special cases.
First, consider the case of multi-user shares single channel.
Since the single-channel optimization problem is reduced from the multi-channel optimization problem, the solution approach of general case can be applied to this case, and the solving algorithm for this case has low computation complexity, which is a combination of analytical and bisection-search methods.
Then, for the case of single-user occupies all channel, the optimization problem is simplified and an optimal solving algorithm with closed-form solutions is proposed.

\end{abstract}


\section{Introduction} \label{s:introduction}

Due to the rapid development of applications, new applications such as speech recognition, virtual reality, image recognition and online games require a great deal of computation. Considering the limited battery lifetime and computation capacity, it is challenging for mobile users (MUs) to accomplish the computation tasks locally.
However, if these tasks are processed centrally by the data center, too much bandwidth and computation resources will be consumed and seriously affect the experience of MUs. Mobile edge computing(MEC) refers to proximal service provided by edge servers rather than the remote data center. Through proximal edge servers, application data can be computed on the vicinity, which greatly reduces the communication load, service latency, and achieve longer battery lives \cite{survey_com2017mao, survey_archi2017mach}. Due to the advantages aforementioned, MEC becomes a promising technology for large-scale wireless networks with massive computation tasks or latency sensitive tasks.

Compared with centralized data centers, MEC providers have relatively limited computation and communication resources. Specially, in multiuser MEC systems, the effective allocation of communication resources for data offloading is a great challenge and multi-access is an efficiently way for MUs to offload their tasks, which can further improves the flexibility of offloading.
Therefore, the main problem faced by the multiuser MEC system is how to allocate limited computation resources effectively to satisfy the offloading demands of MUs and achieve the overall optimal utility.

Since the battery lifetime of MUs are always constrained, another important objective in MEC system is to be energy efficient, i.e., to save energy for MUs within the limitation of latency. In order to better realize the benefits of OFDM in flexible multiple access, adequate communication resource allocation is indispensable.
However, subchannel multiplexing, i.e., each subchannel can be occupied by multiple MUs in multiuser multiaccess MEC system has not yet been studied in the existing works.

In this paper, motivated by above observations, we focus on jointly optimizing the time transmission, local computation speed, task offloading, and subchannel assignment to facilitate the realization of energy-efficient multiuser multiaccess MEC system and utilize the advantages of subchannel multiplexing in OFDMA to reduce the energy consumption in multiuser MEC system.
For general and some special cases, the optimal and low-complexity algorithm are proposed by analyzing the property of the formulated problem.
The main contributions of our work can be summarized as follows:

1) We study an energy-efficient multiuser multi-access partial offloading MEC system where task offloading strategy, radio and computation resource allocation are jointly optimized. An optimization problem is formulated to minimize the weighted sum energy consumption of all MUs within the constraints of communication and radio resources, computation capacity and latency. The formulated optimization problem is nonconvex and is transformed to an equivalent convex optimization problem.

2) For the general multiuser multi-access case, we propose an optimal energy-efficient task offloading and resource allocation strategy, called MMOS algorithm. By exploiting the separable structure for the transformed convex problem, we decouple the problem into two subproblems, corresponding to optimizing the computation capacity allocation and offloading assignment and subchannel bandwidth and proposed an iterative solution method to alternatively solve the subproblems.
More importantly, regarding the distribution of computation capacity among MUs, it is observed that the MEC edge server will allocate the computation capacity to the MU with better channel gains, and there exists a offloading threshold, under which no offloading will occur.

3) To gain more insights, consider the special case where the multiuser share single channel and single user occupies all subchannels. The former case is also transformed and decomposed into two subproblems and similar procedures can be adopted for solving this problem. Meanwhile, the original problem can be simplified as the problem of joint optimization of transmission time and offloading assignment for the latter case. Then, a simple solving algorithm combining analytical and bilevel optimization is deduced for the latter case problem, which has low computation complexity and can achieve global optimal utility.

The rest of paper is organized as follows. In Section \ref{s:main_related_works}, we present the related works. Section \ref{s:main_system model} describes the system model, which is followed by problem formulation and analysis in Section \ref{s:main_problem formation}. In Section \ref{s:main_general}, the optimal energy-efficient task offloading and resource allocation strategy for the general case is proposed, whose convergence and computation complexity is also studied. The optimal and simple solving algorithms for special cases are devised in Section \ref{s:main_special}. Finally, conclusions are provided in Section \ref{s:main_conclusion}.

\section{Related Works} \label{s:main_related_works}
Since computation tasks can be offloaded to the edge server to release the computation burden of MUs, MEC system has been investigated in plenty of pioneer works related with the radio and computational resource allocation.
In general, the basic offloading paradigm can be divided into two categories: binary offloading and partial offloading \cite{survey_archi2017mach, survey_com2017mao}. The former is indivisible continuous processing tasks or relative simple tasks that cannot be partitioned and has to be executed as a whole, while the latter is the task containing multiple procedures and can be divided into several parts for local and edge execution.
Then, the respective works can be explained from the perspective of users and edge servers.

For the perspective of users, with limitation for instance on the battery life and latency requirement of MU, tasks can be binary or partial offloaded to the MEC server. Hence, for the resource-constrained users, the common optimization goals can be summarized as energy consumption \cite{energy_effi, noma1egy_2021ding, noma3egy_2021tsang, ofdma2egy_2019wen}, latency of task completion \cite{latency_opt, tdma1lat_2019xing, noma2lat_2021tsang, ofdma1lat_2020saleem} and a combination of both \cite{enerla_cent1, enerla_cent2, effi_multiu}. 
To achieve these research goals, joint optimization of offloaded data with communication resources\cite{energy_effi, noma1egy_2021ding, noma2lat_2021tsang, ofdma2egy_2019wen}, computing resources \cite{enerla_cent1}, or both \cite{enerla_cent2, tdma1lat_2019xing, noma3egy_2021tsang, ofdma1lat_2020saleem} is performed.


For the perspective of edge servers, its computational resources is also highly constrained due to finite size and configuration\cite{survey_com2017mao}.
Besides, the task processing capabilities of the edge server can be divided into single-task and multi-task parallel types, corresponding to small edge service nodes and resourceful edge servers. Note that multi-core or virtual machine (VM) enables transient customization of MEC computation resource allocation for achieving multi-task parallel processing.
For ease of presentation, the above two types are called as the single-task type and multi-task type.
Thus, it is necessary to design an offloading strategy for single-task type cite and multi-task type cite MEC system with finite radio and computational resources.

For single-task type processing \cite{single1_2017mao}, \cite{single2_2019alame}, the MEC edge server executes the offloaded tasks one by one. In this situation, the authors in \cite{single1_2017mao} jointly optimize task offloading scheduling and transmit power allocation to minimize weighted sum of executing delay and device energy consumption based on single edge server, while in \cite{single2_2019alame}, authors exploited multiple edger servers for multiuser computation offloading, where different MUs can offload their tasks to different edge servers, and the edger servers optimize the task offloading scheduling and application resource allocation.

For multi-task type processing, the MEC edge server process multiple tasks from different MUs simultaneously, while different MUs reuse time/frequency domain resources for computation offloading, which can be summarized as Time Division Multiple Access (TDMA) \cite{energy_effi, tdma1lat_2019xing}, Non-Orthogonal Multiple Access (NOMA) \cite{noma1egy_2021ding, noma2lat_2021tsang, noma3egy_2021tsang} and Orthogonal Frequency-Division Multiple Access (OFDMA) \cite{ofdma1lat_2020saleem, ofdma2egy_2019wen}.
In \cite{energy_effi, tdma1lat_2019xing}, the authors adopt TDMA transmission protocol for MUs to offload their computational tasks over orthogonal predetermined time slots and aim for minimizing weight sum of energy consumption \cite{energy_effi} and latency \cite{tdma1lat_2019xing} for optimizing task assignment, time for offloading, and computation resources. However, the spectrum utilization rate of the TDMA system is low, and the system capacity is small. Therefore, MEC system based on TDMA has an upper bound for serving MUs.
The existing works based on NOMA protocols \cite{noma1egy_2021ding, noma2lat_2021tsang, noma3egy_2021tsang} focus on enable multiple MUs to share the same time-frequency resource to achieve higher spectral efficiency. In NOMA based systems, it is always needs to consider optimizing power allocation, authors minimize the energy consumption \cite{noma1egy_2021ding, noma3egy_2021tsang} with subchannel allocation and task assignment, while minimize the latency \cite{noma2lat_2021tsang} with transmission time, under some deadline constraints. The largest resistance of applying NOMA scheme to the MEC system is the limitation on the number of user groups and the system complexity increases significantly with the increase of user scale, which makes it difficult to implement in general large-scale user scenarios.

Some studies \cite{energy_effi, ofdma1lat_2020saleem, ofdma2egy_2019wen} with OFDMA protocol jointly optimize subchannel assignment, task assignment, communication and computation resources for the minimization of latency \cite{ofdma1lat_2020saleem} or weighted sum of energy consumption \cite{energy_effi, ofdma2egy_2019wen} and subchannel assignment is a significant feature.
Such studies will always transform the objective minimization problem to a mixed integer nonlinear programming (MINLP) problem due to binary subchannel assignment (i.e., subchannel assignment indicators have binary values), which is generally NP-hard and can only obtain the sub-optimal solution of the equivalent transformed problem.
However, the authors in \cite{energy_effi, ofdma1lat_2020saleem, ofdma2egy_2019wen} assume that each subchannel can be occupied by one user at most and each user can be allocated with multiple subchannels, and none of them has taken into account of the subchannel multiplexing, which is able to allocate channels more flexibly to meet the differentiated needs of different users.

Therefore, the study of the optimization on the multiuser multiaccess MEC systems still have not been sufficiently investigated, especially in the respects of subchannel multiplexing and time protocol design, which are expected to be properly solved in this paper.

\section{System Model} \label{s:main_system model}
\begin{figure}
  \centering
  \includegraphics[width=1.5\figwidth]{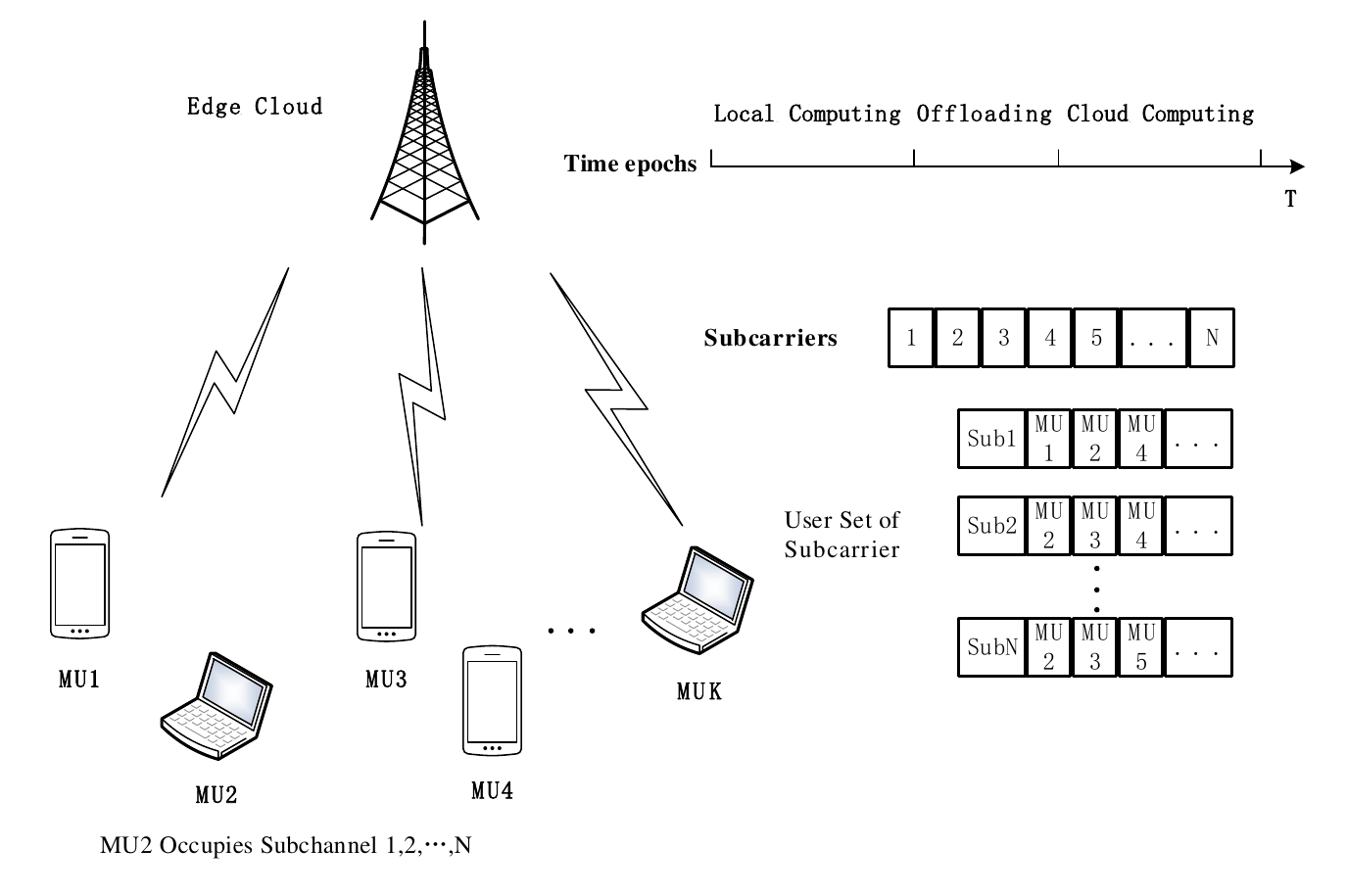}
  \caption{Multiuser multi-access MEC system}\label{f:scene}
\end{figure}

As shown in Fig. \ref{f:scene},
a MEC system is considered with $\mathcal{N}$ frequency bands of single edge server (termed subchannels in the sequel) and $\mathcal{K}$ MUs, which constitutes the set $\mathcal{N}\triangleq \{1,2,...,N\}$ and $\mathcal{K}\triangleq \{1,2,...,K\}$, respectively.
Each MU has a computational task which can be completed with the help of the edge server, where part or whole of the task can be offloaded to the edge server to satisfy the latency requirement, and other part of the task is finished locally.
Considering the computational task executed by $k$th MU is latency-critical, its input bits should be completed within latency constraint, i.e., $T_k$ (seconds), and we assume that $T_k$ is smaller than the channel coherence time, which means the channel gain remains unchanged during the MU completes its task.

\begin{figure}
  \centering
  \includegraphics[width=1.5\figwidth]{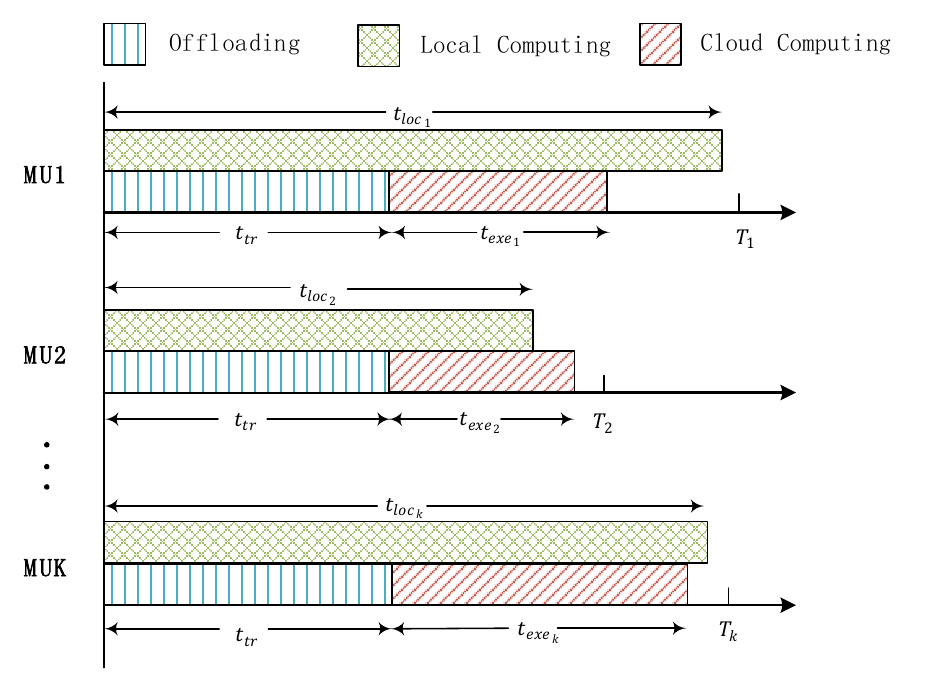}
  \caption{The frame structure for the proposed MEC protocol}\label{f:frame_structure}
\end{figure}

In this paper, we consider a three-phase time protocol, whose frame structure can be shown as in Fig. \ref{f:frame_structure}.
The $k$th MU first determines the amount of data for offloading in the preparation phase, and it is divided into local and offloaded parts for future processing. Then, the local part starts the execution phase and computes the tasks. For the offloaded part, the $k$ MU transmits part of the data to the edge server by subchannel $n$, $n \in \mathcal{N}$ in the offloading phase, and the edge server executes the offloaded tasks and feedbacks the computation results to the associated MU in the execution phase.

Note that after the preparation phase, local and offload part start the following processing simultaneously, which can be achieved through full duplex. Moreover, at each time slot during the offloading phase, the local MU can offloads tasks via multiple subchannels simultaneously and each subchannel can be occupied by multiple MUs.
Next, we introduce the three-phase protocol in detail.

\subsection{Local Computation Model} \label{s:system_local}
For the computational task of $k$th MU, $k\in \mathcal{K}$, the amount of data for computing  $R_k$ (in unit of bits) can be arbitrarily partitioned into two parts: offloading and local computing, which can be denoted by $l_k \geq 0$ and $(R_k - l_k) \geq 0$, respectively.

The total computation amount of $k$th MU is $C_k R_k$ for $k\in \mathcal{K}$, in which $C_k$ denotes the amount of computation (in unit of CPU cycles) for computing one bit data of the task.

We model the local power consumption of CPU as $p = \varepsilon f^3$ \cite{MEC_dynamic}, where $\varepsilon$ and $f$ are the coefficient depending on chip architecture and computation speed , respectively.
Since the computation energy consumption of the MU can be optimized by adjusting computation speed through dynamic voltage scaling (DVS) technology, we suppose the computation speed of $k$th MU is $f_{l_k}$, and its upper bound is $F_k$.
Then, the local execution time $t_{loc_k}$ and local energy consumption $E_{loc_k}$ can be given as
\begin{align}
  t_{loc_k} & = \frac{C_k (R_k - l_k)}{f_{l_k}}, \label{e:tloc} \\
  E_{loc_k} & = \varepsilon_k C_k (R_k - l_k) f_{l_k}^2. \label{e:Eloc}
\end{align}
\subsection{Offloading Model} \label{s:system_offloading}
The MU offload data to the edge server through wireless transmission. Suppose the $\mathcal{N}$ subchannels have equal bandwidth $\tau$ and each subchannel can be assigned to different MUs. In other words, the $k$th MU has an assigned channel with bandwidth $\tau_{k,n}$ from $n$th subchannel.

Besides, we denote $h_{k,n}$ as the channel gain between the edge server and the $n$th MU on subchannel $n$.
Since the assigned subchannel bandwidth is only a small part of system bandwidth, it is reasonable to assume that the subchannel bandwidth assigned for the MU is smaller than the coherence bandwidth, and the channel gain is flat within the bandwidth, which means it remains unchange during offloading.

We denote $P_{k,n}$ as the transmission power of $k$th MU on subchannel $n$, and normalize the PSD of background noise as $1$. Then, the transmission rate of $k$th MU on subcarrier $n$ can be given as $r_{k,n} = \tau_{k,n} ln \left( 1+ \frac{P_{k,n} h_{k,n}}{\tau_{k,n}}\right)$.

Denote $l_{k,n}$ as the amount of offloading data of $k$th MU on subchannel $n$, then we have $l_k = \sum_{n=1}^{N} l_{k,n}$.
Considering all MUs offload data with same transmission time $t_{tr}$, $l_{k,n}$ can be written as $l_{k,n} = r_{k,n} t_{tr}$.
With previous definitions, $P_{k,n}$ can be expressed as $P_{k,n} = \left( e^{\frac{l_{k,n}}{\tau_{k,n} t_{tr}}} -1 \right) \frac{\tau_{k,n}}{h_{k,n}}$, and the transmission energy consumption for offloading data of $k$th MU can be given as
\begin{equation}\label{e:Eoff}
  E_{\text{off}_k} = \sum_{n=1}^{N} P_{k,n} t_{tr} = \sum_{n=1}^{N} (e^{\frac{l_{k,n}}{\tau_{k,n} t_{tr}}} -1) \frac{\tau_{k,n} t_{tr}}{h_{k,n}}.
\end{equation}

Note that, for $l_{k,n}$ and $\tau_{k,n}$, only the following two cases will happen for $\forall k \in \mathcal{K}, \forall n \in \mathcal{N}$: ``$l_{k,n} = 0, \tau_{k,n} = 0$'' and ``$l_{k,n} > 0, \tau_{k,n} > 0$''.
The other two cases ``$l_{k,n} > 0, \tau_{k,n} = 0$'' and ``$l_{k,n} = 0, \tau_{k,n} > 0$'' will not happen under practical conditions. Once ``$l_{k,n} > 0, \tau_{k,n} = 0$'' (or ``$l_{k,n} = 0, \tau_{k,n} > 0$'') happens, it is better to allocate the nonzero $l_{k,n}$ (or $\tau_{k,n}$) to the subchannel whose bandwidth (or offloading data size) is nonzero.

\subsection{Edge Computation Model} \label{s:system_edge}
For edge computation, the edge server assigns its computation capacity $f_{c_k}$ (in unit of CPU cycles/second) to the $k$th MU, whose total computation capacity is $F_c$ (also in unit of CPU cycles/second), and we have $\sum_{k=1}^{K} f_{c_k} \leq F_c$. Here, $f_{c_k}$ is fixed for the duration of task execution.
Then, the corresponding execution time $t_{\text{exe}_k}$ can be given as
\begin{equation}\label{e:texe}
  t_{\text{exe}_k} = \frac{C_k l_k}{f_{c_k}} = \frac{C_k \sum_{n=1}^{N} l_{k,n}}{f_{c_k}}
\end{equation}

Thus, for $k$th MU, the consumed time $t_{s_k}$ for computing through edge server can be expressed as
\begin{equation}\label{e:tserver}
  t_{s_k} = t_{tr} + t_{\text{exe}_k}.
\end{equation}
Since the computation result of the offloaded data is much less and the backhaul transmission rate is much larger \cite{effi_multiu}, the backhaul transmission time could be ignored.
Therefore, the total task processing latency for $k$th MU can be expressed as $\max{ \{t_{loc_k}, t_{s_k} \}}$.

Besides, the energy consumption for $k$th MU $E_k(t_{tr}, f_{l_k}, f_{c_k}, l_{k,n}, \tau_{k,n})$ can be expressed as
\begin{equation}\label{e:E}
  E_k(t_{tr}, f_{l_k}, f_{c_k}, l_{k,n}, \tau_{k,n}) = E_{loc_k} + E_{\text{off}_k}
\end{equation}
Due to the fact that the energy consumption of offloading and result reception are of different orders of magnitude, hence, we ignore the energy consumption for result reception of MUs \cite{multiu_min}.

\section{Problem Formulation and Analysis} \label{s:main_problem formation}
\subsection{Problem Formulation} \label{s:problem_formation}
Considering transmission time $t_{tr}$, local computational speeds $\bm{f_l} = \{f_{l_k}\}$, assigned edge computational speeds $\bm{f_c} = \{f_{c_k}\}$, offloading data size $\bm{l}= \{l_{k,n}\}$ and subchannel bandwidth allocation $\bm{\tau} = \{\tau_{k,n}\}$ for $\forall k\in \mathcal{K}, \forall n \in \mathcal{N}$, we formulate the multi-user multi-access partial offloading problem which minimizes the sum of MUs' energy consumption as follows:
\begin{prob} \label{p:problem}
\begin{subequations}
\begin{align}
  \mathop{\min} \limits_{ t_{tr},\bm{ f_l, f_c, l, \tau }} \quad & \sum_{k=1}^{K} \omega_k E_k(t_{tr}, f_{l_k}, f_{c_k}, l_{k,n}, b_{k,n})   \nonumber\\
  \text{s.t.}  \quad & \max{ \{t_{loc_k}, t_{s_k} \}} \leq T_k, \forall k \in \mathcal{K}, \label{e:stTk}\\
               \quad & 0 \leq f_{l_k} \leq F_k, \forall k \in \mathcal{K}, \label{e:stFk}\\
               \quad & \sum_{k=1}^{K} f_{c_k} \leq F_c, \label{e:stFc} \\
               \quad & f_{c_k} \geq 0, \forall k \in \mathcal{K}, \label{e:stfc}\\
               \quad & \sum_{n=1}^{N}l_{k,n} \leq R_k, \forall k \in \mathcal{K},  \label{e:stRk}\\
               \quad & l_{k,n} \geq 0, \forall k \in \mathcal{K}, \forall n \in \mathcal{N} \label{e:stlkn}\\
               \quad & \sum_{k=1}^{K} \tau_{k,n} \leq \tau, \forall n \in \mathcal{N}, \label{e:sttau}\\
               \quad & \tau_{k,n} \geq 0, \forall k \in \mathcal{K}, \forall n \in \mathcal{N}, \label{e:sttau2}
\end{align}
\end{subequations}
\end{prob}
where $E_k(t_{tr}, f_{l_k}, f_{c_k}, l_{k,n}, \tau_{k,n})$ can be expressed as

$E_k(t_{tr}, f_{l_k}, f_{c_k}, l_{k,n}, \tau_{k,n}) = \varepsilon_k C_k (R_k - \sum_{n=1}^{N}l_{k,n}) f_{l_k}^2 + \sum_{n=1}^{N} (e^{\frac{l_{k,n}}{\tau_{k,n} t_{tr}}} -1)\frac{\tau_{k,n} t_{tr}}{h_{k,n}}$.

Besides, $\omega_k$ is the weighting factor for $k$th MU, which reflects the urgency for energy saving, namely, the larger the $\omega_k$, the less the remaining energy.

In Problem \ref{p:problem}, we aim at minimizing weighted sum of energy consumption for MUs subjects to constraitns (\ref{e:stTk})-(\ref{e:stlkn}), which reflects a tradeoff of energy consumption among MUs. (\ref{e:stTk}) reflects latency requirements of MUs; (\ref{e:stFk}) and (\ref{e:stFc}) represent the maximum computation capacity constraint for each MU and edge server, respectively; (\ref{e:stfc}) indicates that the assigned computation resource from edge server is non-negative; (\ref{e:stRk}) and (\ref{e:stlkn}) stand for the total offloaded data constraint for $k$th MU and the non-negativity of the offloaded data in each subchannel; (\ref{e:sttau}) and (\ref{e:sttau2}) state that the allocated subchannel bandwidth to MUs can not exceed the total bandwidth of subchannel and its non-negative.


\subsection{Problem Analysis} \label{s:problem_analysis}
\textit{1) Feasibility:}
In order to guarantee the feasibility of Problem \ref{p:problem}, we study its feasible region. According to constraints (\ref{e:stTk}), (\ref{e:stFk}), (\ref{e:stFc}), (\ref{e:stRk}) and (\ref{e:stlkn}), we have
\begin{equation}\label{e:feasiblelk}
  l_k \in \left[l_{k_{bd}}^-, l_{k_{bd}}^+ \right]
\end{equation}
where $l_{k_{bd}}^-$ and $l_{k_{bd}}^+$ are defined as $l_{k_{bd}}^- = \max \{R_k - \frac{T_k F_k}{C_k}, 0\}$ and $l_{k_{bd}}^+ = \min \{\frac{(T_k - t_{tr}) F_c}{C_k}, R_k\}$, respectively.

To make Equ.(\ref{e:feasiblelk}) established, $R_k - \frac{T_k F_k}{C_k} \leq \frac{(T_k - t_{tr}) F_c}{C_k}$ should hold, thus the condition for non-empty feasible set of $l_k$ is $T_k \geq T_k^{\text{Feas}} \triangleq \frac{R_k C_k + t_{tr} F_c}{F_k + F_c}$, which is the feasible region of $T_k$ for Problem \ref{p:problem}.
Besides, the condition for full offloading can be deduced as $T_k \geq T_k^{\text{Off}} \triangleq \frac{R_k C_k}{F_c} + t_{tr}$, and $T_k^{\text{Off}}$ is larger than $T_k^{\text{Feas}}$ obviously.
Therefore, the only partial offloading case is $T_k \in \left[T_k^{\text{Feas}}, T_k^{\text{Off}} \right)$.

Hence, we assume that the system parameters satisfy the feasibility of Problem \ref{p:problem}, as considered in many works \cite{effi_multiu, multiu_min, energy_effi, enerla_cent2}.

\textit{2) Convexity:}

Since the formulated problem is nondeterministic and the coupled variables make the problem more intractable, it is hard to solve the problem directly.
Motivated by the difficulty of solving the problem, we decouple the variables and transform the formulated problem into an equivalent convex optimization problem, and design a low-complexity algorithm  to find the optimal solution.

To address Problem \ref{p:problem}, we first study the optimal task offloading and resource allocation strategy for the general case that there are multiple MUs and multiple channels (e.g., $|\mathcal{K}|>1$, $|\mathcal{N}|>1$) by adopting analytical and block coordinate decent (BCD) methods \cite{xu2012block}.
Subsequently, we deduce more insightful structures of the optimal strategy for two special cases, thus multiple users share single channel ($|\mathcal{K}|>1$, $|\mathcal{N}|=1$) and single user occupies all channel ($|\mathcal{K}|=1$, $|\mathcal{N}|>1$).

\section{Optimal Solution in General Case} \label{s:main_general}

In this section, we consider the general case of multiuser and multichannel and study its energy-efficient task offloading and resource allocation strategy.
The optimal solution strategy can be summarized as follows:
First, we deduce the optimal local computation speed and transmission time, and then transform Problem \ref{p:problem} into Problem \ref{p:problem_flk}, which is an equivalent convex problem.
To characterize the structure of the optimal strategy, we decompose Problem \ref{p:problem_flk} into two subproblems and propose a block coordinate descent algorithm to solve Problem \ref{p:problem_tau} and Problem \ref{p:problem_lknfck} alternatively. Specifically, we solve Problem \ref{p:problem_tau} with  analytical and bisection search method for subchannel bandwidth allocation.
For Problem \ref{p:problem_lknfck}, two different cases are identified depending on whether the Lagrange multiplier associated with the constraint is equal or larger than zero. The solution approach is derived to achieve the minimal utility for both cases and the solution corresponding to the smaller minimal utility is the optimal solution for task offloading and computation resource allocation.

\subsection{Optimal Local Resource Allocation Strategy} \label{s:general_local}
This subsection aims at finding the optimize local computation speed and transmission time.
Since the local computation speed $f_{l_k}$ is not coupled with other variables, we can first optimize $f_{l_k}$ to simplify Problem \ref{p:problem}.
According to constraint (\ref{e:stTk}), it can be easily derived that $f_{l_k} \geq \frac{C_k (R_k - \sum_{n=1}^{N}l_{k,n})}{T_k}$. Note that $E_k(t_{tr}, f_{l_k}, f_{c_k}, l_{k,n}, \tau_{k,n})$ is monotonically increasing with respect to $f_{l_k}$.
The optimal computation speed $f_{l_k}^*$ for $k \in \mathcal{K}$ of Problem \ref{p:problem} can be expressed as:
\begin{equation}\label{e:op_flk}
  f_{l_k}^* = \frac{C_k (R_k - \sum_{n=1}^{N}l_{k,n})}{T_k}.
\end{equation}

Substituting the optimal $f_{l_k}^*$ into Problem \ref{p:problem}, the origin problem can be simplified as:
\begin{prob} \label{p:problem_flk}
\begin{subequations}
\begin{align}
  \mathop{\min} \limits_{t_{tr},\bm{ f_c, l, \tau }} \quad &  \sum_{k=1}^{K} \omega_k E_k(t_{tr}, f_{c_k}, l_{k,n}, \tau_{k,n})  \nonumber\\
  \text{s.t.}  \quad & \frac{C_k \sum_{n=1}^{N}l_{k,n}}{f_{c_k}} \leq T_k - t_{tr}, \forall k \in \mathcal{K}, \label{e:st2Tk}\\
               \quad & \text{Constraint}~(\ref{e:stFc}) \text{~--~Constraint} ~(\ref{e:sttau2}), \nonumber
\end{align}
\end{subequations}
\end{prob}
where $E_k(t_{tr}, f_{c_k}, l_{k,n}, \tau_{k,n})$ can be expressed as
\begin{equation}\label{e:E4}
\begin{array}{ll}
  E_k(t_{tr}, f_{c_k}, l_{k,n}, \tau_{k,n}) = 
   \frac{\varepsilon_k C_k^3 (R_k - \sum_{n=1}^{N}l_{k,n})^3}{T_k^2} +
  \sum_{n=1}^{N} (e^{\frac{l_{k,n}}{\tau_{k,n} t_{tr}}} -1)\frac{\tau_{k,n} t_{tr}}{h_{k,n}}.
\end{array}
\end{equation}
Besides, (\ref{e:st2Tk}) is evolved from (\ref{e:stTk}). Note that Problem \ref{p:problem_flk} is still nondeterministic, which is hard to solve directly.
Then, we transform the problem into a series of solvable convex subproblems, and obtain the optimal solution of Problem \ref{p:problem_flk} by solving the subproblems.

For function $E_k(t_{tr}, f_{c_k}, l_{k,n}, \tau_{k,n})$, there is the following property.
\begin{lem} \label{lem:P2_tr}
$E_k(t_{tr}, f_{c_k}, l_{k,n}, \tau_{k,n})$ is a monotonically decreasing function with $t_{tr}$ for $t_{tr} \geq 0$, and $\forall k \in \mathcal{K}$,  $\forall n \in \mathcal{N}$.
\end{lem}
\begin{IEEEproof}
Define $g(t_{tr}) \triangleq (e^{\frac{l_{k,n}}{\tau_{k,n} t_{tr}}} -1)\frac{\tau_{k,n} t_{tr}}{h_{k,n}}$, investigate the first-order and second-order partial derivatives of $g(t_{tr})$, there are $\frac{\partial {g(t_{tr})}}{\partial{t_{tr}}} = \frac{\tau}{h_{k,n}} [e^{\frac{l_{k,n}}{\tau t_{tr}}}(1-\frac{l_{k,n}}{\tau t_{tr}}) -1]$, and $\frac{\partial^2 g(t_{tr})}{\partial t_{tr}^2} =  \frac{l_{k,n}^2 e^{\frac{l_{k,n}}{\tau t_{tr}}}}{h_{k,n} \tau t_{tr}^3}$. When $l_{k,n} \geq 0$ and $t_{tr} \geq 0$, there always exists $\frac{\partial^2 g(t_{tr})}{\partial t_{tr}^2} \geq 0$, which indicates $\frac{\partial {g(t_{tr})}}{\partial{t_{tr}}}$ keep on increasing with $t_{tr}$.
In addition, $\lim \limits_{t_{tr} \to 0^+} \frac{\partial {g(t_{tr})}}{\partial{t_{tr}}} = -\infty$ and $\lim \limits_{t_{tr} \to \infty} \frac{\partial {g(t_{tr})}}{\partial{t_{tr}}} = 0$. Hence, $\frac{\partial {g(t_{tr})}}{\partial{t_{tr}}}$ is always less than $0$ for $t_{tr} \geq 0$, which proves the decreasing monotonicity of $g(t_{tr})$ with $t_{tr}$.
\end{IEEEproof}

Since $E_k(t_{tr}, f_{c_k}, l_{k,n}, \tau_{k,n})$ is monotonic with respect to $t_{tr}$, the optimal transmission time $t_{tr}^*$ can be found by bisection search, and the optimal solution of Problem \ref{p:problem_flk} can be achieved at the optimal transmission time $t_{tr}^*$.

With the given $t_{tr}^{(r)}$ in $r$-th round, it can be easily proved that $E_k(t_{tr}^{(r)}, f_{c_k}, l_{k,n}, \tau_{k,n})$ is jointly convex with respect to $\{f_{c_k}, l_{k,n}, \tau_{k,n}\}$ for $\forall k \in \mathcal{K}$. Moreover, the left-hand functions of constraints of Problem \ref{p:problem_flk} are all convex with respect to $\{f_{c_k}, l_{k,n}, \tau_{k,n}\}$, $\forall k \in \mathcal{K}$, $\forall n \in \mathcal{N}$.
Therefore, Problem \ref{p:problem_flk} is a convex problem over $\{f_{c_k}, l_{k,n}, \tau_{k,n}\}$ when $t_{tr}$ fixed, which is also block multi-convex.

Hence, for the given $t_{tr}^*$, the optimal solution of Problem \ref{p:problem_flk} can be obtained by applying {\it block coordinate decent} (BCD) method with alternately solving the following two subproblems:


\begin{prob} \label{p:problem_tau}
\begin{subequations}
\begin{align}
  \mathop{\min} \limits_{\bm{ \tau }} \quad &  \sum_{k=1}^{K} \omega_k E_k( \overline{f_{c_k}}, \overline{l_{k,n}}, \tau_{k,n})  \nonumber\\
  \text{s.t.}  \quad & \text{Constraint}~(\ref{e:sttau}), ~(\ref{e:sttau2}). \nonumber
\end{align}
\end{subequations}
\end{prob}

\begin{prob} \label{p:problem_lknfck}
\begin{subequations}
\begin{align}
  \mathop{\min} \limits_{\bm{ f_c, l }} \quad &  \sum_{k=1}^{K} \omega_k E_k(f_{c_k}, l_{k,n}, \overline{\tau_{k,n}})  \nonumber\\
  \text{s.t.}  \quad & \text{Constraints}~(\ref{e:stFc}) \text{--} (\ref{e:stlkn}), ~(\ref{e:st2Tk}). \nonumber
\end{align}
\end{subequations}
\end{prob}
Here, Problem \ref{p:problem_tau} optimizes subchannel bandwidth $\tau_{k,n}$ for the given $\overline{f_{c_k}}$, and $\overline{l_{k,n}}$, while Problem \ref{p:problem_lknfck} optimizes computation capacity allocation and offloading management for the given $\overline{\tau_{k,n}}$.

\subsection{Optimal Transmission Resource Allocation Strategy} \label{s:general_channel}
In this subsection, we aim at finding the optimal subchannel bandwidth $\tau_{k,n}$ with given computation capacity allocation and offloading partition.
Observed that Problem \ref{p:problem_tau} is convex with respect to $\tau_{k,n}$, which can be solved by numerical optimization methods with numerical optimal solution. In this subsection, we adopt Karush-Kuhn-Tucker (KKT) conditions to analyze the special property of the problem and achieve closed-form optimal solution.

Since Problem \ref{p:problem_tau} satisfies Slater's condition, the KKT conditions serves as a sufficient and necessary condition for the optimal solution can be listed as follows \cite{Boyd}.
\begin{subequations}
\begin{align}
   -\frac{t_{tr}^*}{h_{k,n}} \left[ e^{\frac{l_{k,n}}{\tau_{k,n} t_{tr}^*}} (\frac{l_{k,n}}{\tau_{k,n} t_{tr}^*} -1) +1 \right] + \lambda_n - v_{k,n} &= 0, \forall k \in \mathcal{K}, n \in \mathcal{N} \label{e:tau_kkt1}\\
   \lambda_n (\sum_{k=1}^{K} \tau_{k,n} - \tau) &= 0, \forall n \in \mathcal{N} \label{e:tau_kkt2}\\
   v_{k,n} \tau_{k,n} &= 0, \forall k \in \mathcal{K}, \forall n \in \mathcal{N} \label{e:tau_kkt3}\\
   \text{Constraints}~(\ref{e:sttau}),& ~(\ref{e:sttau2}). \nonumber
\end{align}
\end{subequations}
in which $\lambda_n$ and $v_{k,n}$ are non-negative Lagrange multipliers associated with constraints (\ref{e:sttau}) and (\ref{e:sttau2}) respectively.

With mathematical operations, the optimal policy for subchannel bandwidth allocation $\forall k \in \mathcal{K}, \forall n \in \mathcal{N}$ can be expressed as
\begin{equation}\label{e:opt_tau}
\tau_{k,n}^* =
\begin{cases}
\frac{l_{k,n}}{t_{tr}^* \left( W_{k,n} + 1 \right)}, & W_{k,n} > -1, \\
  0, & \mbox{otherwise}.
\end{cases}
\end{equation}
where $W_{k,n} = W_0 \left(- \frac{1}{e} [\frac{h_{k,n}}{t_{tr}^*} (v_{k,n} - \lambda_n) +1]\right)$, and $W_0(\cdot)$ denotes the $0$th branch of the Lambert W function. 
Note that, when $W_{k,n} = -1$, there is $l_{k,n} = 0$, and the corresponding $\tau_{k,n}$ should be $0$, which means that the $k$th MU does not select $n$th subchannel to offload, while the edge server does not allocate bandwidth of $n$th subchannel to $k$th MU.
For the $k$th MU, define the active subchannel set as $\mathcal{A}_k \triangleq \{ i| l_{k,i}>0, \tau_{k,i}>0, i \in \mathcal{N}\}$, which means the $k$th MU offloads on $i$th subchannel and bandwidth of $i$th subchannel is given to $k$th MU.

To obtain the optimal subchannel bandwidth $\tau_{k,n}^*$ via Equ.(\ref{e:opt_tau}) requires computing the Lagrange multipliers $\lambda_n$ and $v_{k,n}$ by solving (\ref{e:tau_kkt2}) and (\ref{e:tau_kkt3}).
For the KKT condition, the following lemma can be obtained.
\begin{lem} \label{lem:gen_lkn_lambda}
For $\forall n \in \mathcal{N}$, $\lambda_n>0$.
\end{lem}
\begin{IEEEproof}
For $\forall k \in \mathcal{A}_k$, $v_{k,n}=0$ according to constraint (\ref{e:tau_kkt3}), and constraint (\ref{e:tau_kkt1}) can be rewritten as
\begin{equation}\label{e:spec_tau_lamda}
  e^{\frac{l_{k,n}}{\tau_{k,n} t_{tr}^*}} (\frac{l_{k,n}}{\tau_{k,n} t_{tr}^*} -1) = \lambda_n \frac{h_{k,n}}{t_{tr}^*} -1, \forall k \in \mathcal{K}, \forall n \in \mathcal{N}
\end{equation}
Define $S(\tau_{k,n}) \triangleq e^{\frac{l_{k,n}}{\tau_{k,n} t_{tr}^*}} (\frac{l_{k,n}}{\tau_{k,n} t_{tr}^*} -1)$, which is a monotonic decreasing function of nonnegative $\tau_{k,n}$.
For $\forall n \in \mathcal{A}_k$, $\lim \limits_{\tau_{k,n} \to \infty} S(\tau_{k,n})=-1$, thus, $S(\tau_{k,n})>-1$ for $\tau_{k,n} \in [0, \tau]$, which illustrates $\lambda_n>0$.

For $\forall n \notin \mathcal{A}_k$, it can be seen that $l_{k,n}=0$ and $\tau_{k,n}=0$, and constraint (\ref{e:tau_kkt1}) can be simplified as $\lambda_n - v_{k,n}=0$, i.e., $\lambda_n= v_{k,n}>0$.

In summary, $\lambda_n>0$ in constraint (\ref{e:tau_kkt2}) for $\forall n \in \mathcal{N}$, which means constraint (\ref{e:sttau}) is active, namely the bandwidth of subchannel $n$ is fully allocated by the MUs.
Here, we define the constraint to be active when the equality in an inequality holds, and inactive denotes the inequality in an inequality holds \cite{robust2016fan}.
\end{IEEEproof}

From Lemma \ref{lem:gen_lkn_lambda}, we can know that constraint (\ref{e:sttau}) is active, i.e., $\sum_{k=1}^{K} \tau_{k,n} = \tau, \forall n \in \mathcal{N}$. Substituting the optimal subchannel bandwidth $\tau_{k,n}^*$ from Equ.(\ref{e:opt_tau}) into above equation, we can get an equation about $\lambda_n$, which can be efficiently found via bisection search  \cite{Boyd}.
With the solved $\lambda_n$, the optimal subchannel bandwidth $\{\tau_{k,n}\}$ can be obtained from Equ.(\ref{e:opt_tau}).

\subsection{Optimal Offloading and Edge Computation Resource Allocation Strategy} \label{s:general_offloading}
For given subchannel bandwidth $\{\tau_{k,n}\}$, this subsection focus on optimizing the computation capacity allocation $\{f_{c_k}\}$ and offloading partition $\{l_{k,n}\}$ for solving Problem \ref{p:problem_lknfck}.
Considering practical factors, the allocation of computation capacity of server just satisfy the latency constraint of the MU. Besides, we know $f_{c_k} \geq \frac{C_k \sum_{n=1}^{N} l_{k,n}}{T_k - t_{tr}}$ from constraint (\ref{e:st2Tk}), so it can be deduced that
\begin{equation}\label{e:opt_fck}
  f_{c_k} = \frac{C_k \sum_{n=1}^{N} l_{k,n}}{T_k - t_{tr}^*}.
\end{equation}

Moreover, with the optimal subchannel bandwidth $\tau_{k,n}^*$ in Section \ref{s:general_channel}, Problem \ref{p:problem_lknfck} can be rewritten with mathematical transformations as
\begin{prob} \label{p:problem_lknfck3}
\begin{subequations}
\begin{align}
  \mathop{\min} \limits_{\bm{ l }} \quad &  \sum_{k=1}^{K} \omega_k E_k(l_{k,n})  \nonumber\\
  \text{s.t.}  \quad & \sum_{k=1}^{K} \frac{C_k \sum_{n=1}^{N} l_{k,n}}{T_k - t_{tr}^*} \leq F_c, \label{e:Elkn_Fc}\\
               \quad & \text{Constraints} ~(\ref{e:stRk}), ~(\ref{e:stlkn}). \nonumber
\end{align}
\end{subequations}
\end{prob}
where $E_k(l_{k,n}) = \frac{\varepsilon_k C_k^3 (R_k - \sum_{n=1}^{N}l_{k,n})^3}{T_k^2} +
  \sum_{n=1}^{N} e^{\frac{l_{k,n}}{\tau_{k,n}^* t_{tr}^*}} \frac{\tau_{k,n}^* t_{tr}^*}{h_{k,n}}$, $\forall k \in \mathcal{K}, \forall n \in \mathcal{N}$.

Observed that Problem \ref{p:problem_lknfck3} is a convex problem for $l_{k,n}$ and satisfies Slater's condition, its KKT conditions serve as a sufficient and necessary condition for the optimal solution, which can be listed as follows.
\begin{subequations}
\begin{align}
   -\frac{3 \varepsilon_k C_k^3 (R_k - \sum_{n=1}^{N} l_{k,n})^2}{T_k^2} + \frac{1}{h_{k,n}} e^{\frac{l_{k,n}}{\tau_{k,n}^* t_{tr}^*}} + \frac{\Gamma C_k}{T_k - t_{tr}^*} 
   + \delta_k - \zeta_{k,n} &= 0, \forall k \in \mathcal{K}, \forall n \in \mathcal{N}  \label{e:lkn_kkt11}\\
   \Gamma (\sum_{k=1}^{K} \frac{C_k \sum_{n=1}^{N} l_{k,n}}{T_k - t_{tr}^*} - F_c) &=0 \label{e:lkn_kkt44}\\
   \delta_k (\sum_{n=1}^{N} l_{k,n} - R_k) &= 0, \forall k \in \mathcal{K} \label{e:lkn_kkt22}\\
   \zeta_{k,n} l_{k,n} &= 0, \forall k \in \mathcal{K}, \forall n \in \mathcal{N} \label{e:lkn_kkt33}\\
   \text{Constraints}~(\ref{e:stRk}), ~(\ref{e:stlkn}) ,& ~(\ref{e:Elkn_Fc}). \nonumber
\end{align}
\end{subequations}
where $\Gamma$, $\delta_k$ and $\zeta_{k,n}$ are non-negative Lagrange multipliers associated with constraints (\ref{e:stRk}), (\ref{e:stlkn}) and (\ref{e:Elkn_Fc}) respectively.

Define $\Delta_{k,n} \triangleq \frac{3 \varepsilon_k C_k^3 (R_k - \sum_{n=1}^{N} l_{k,n})^2}{T_k^2} - \frac{\Gamma C_k}{T_k - t_{tr}^*} - \delta_k + \zeta_{k,n}, \forall k \in \mathcal{K}, \forall n \in \mathcal{N}$, thus we can obtain $\Delta_{k,n} = \frac{1}{h_{k,n}} e^{\frac{l_{k,n}}{\tau_{k,n}^* t_{tr}^*}}, \forall k \in \mathcal{K}, \forall n \in \mathcal{N}$ from constraint (\ref{e:lkn_kkt11}).
Then, we can derive the offloading data size of $k$th MU as
\begin{equation}\label{e:gen_lkn2}
  l_{k,n} =  \tau_{k,n} t_{tr}^* ln  (\Delta_{k,n} h_{k,n}), \forall k \in \mathcal{K}, \forall n \in \mathcal{N}
\end{equation}
from which, we can know that $l_{k,n}$ is monotonically increasing with respect to $\Delta_{k,n}$.
Therefore, we can transform constraint (\ref{e:lkn_kkt11}) into a function of $\Delta_{k,n}$ as follows.
\begin{equation}\label{e:gen_lkn_delta}
  \frac{\Gamma C_k}{T_k - t_{tr}^*} + \delta_k = \frac{3 \varepsilon_k C_k^3 (R_k - \sum_{n=1}^{N} l_{k,n})^2}{T_k^2}  + \zeta_{k,n} - \Delta_{k,n}, \forall k \in \mathcal{K}, \forall n \in \mathcal{N}
\end{equation}
The right-hand side of Equ.(\ref{e:gen_lkn_delta}) is monotonically decreasing with respect to $\Delta_{k,n}$. Then, we need to solve for the $\Gamma$ and $\delta_k$ to find the optimal solution of $l_{k,n}$.

As Lagrange multipliers, $\Gamma$ and $\delta_k$ should be equal to or larger than zero, which means the corresponding constraint (\ref{e:stFc}) and (\ref{e:stRk}) is uncertain or active.
Next, we first study the property of $\delta_k$ and then discuss $\Gamma$. First, the following lemma shows that $\delta_k=0$ for the $k$th MU.
\begin{lem} \label{lem:lkn_delta2}
For $\forall k \in \mathcal{K}$, $\delta_k=0$.
\end{lem}
\begin{IEEEproof}
For $\forall k \in \mathcal{K}$, when $\forall n \in \mathcal{A}_k$ and $\mathcal{A}_k \neq \emptyset$, $\zeta_{k,n}$ is $0$ according to constraint (\ref{e:lkn_kkt33}), and Equ.(\ref{e:gen_lkn_delta}) can be rewritten as
\begin{equation}\label{e:lkn_delta2}
  \frac{\Gamma C_k}{T_k - t_{tr}^*} + \delta_k = \frac{3 \varepsilon_k C_k^3 (R_k - \sum_{n=1}^{N} l_{k,n})^2}{T_k^2} - \Delta_{k,n}, \forall k \in \mathcal{K}, \forall n \in \mathcal{N}
\end{equation}
Note that $\Delta_{k,n}>0$ for $n \in \mathcal{N}$ and $T_k - t_{tr}^*>0$ according to its definition.

Then, we use proof by contradiction.
For $k \in \mathcal{K}$, when $\delta_k>0$, the constraint (\ref{e:stRk}) is active according to constraint (\ref{e:lkn_kkt22}), thus $\sum_{n=1}^{N} l_{k,n} = R_k$.
Suppose $\delta_k>0$ for a channel belonging to the active channel set, namely $n \in \mathcal{A}_k$,
then the left-hand side of Equ.(\ref{e:lkn_delta2}) is larger than $0$ while the right-hand side is less than $0$. This formulates the contradiction, i.e., for $\forall k \in \mathcal{K}$ and $\forall n \in \mathcal{A}_k$, $\delta_k=0$.
Moreover, for a specific $k$, $\delta_k$ is unchanged for $\forall n \in \mathcal{N}$ according to constraint (\ref{e:lkn_kkt22}), which means $\delta_k=0$ for the $k$th MU for $\forall n \in \mathcal{N}$ when $\mathcal{A}_k \neq \emptyset$.

When $\mathcal{A}_k = \emptyset$ for $\forall k \in \mathcal{K}$, we can know that $l_{k,n} =0$ for $\forall n \in \mathcal{N}$, thus $\sum_{n=1}^{N} l_{k,n}=0$, and derive that the constraint (\ref{e:stRk}) is inactive according to (\ref{e:lkn_kkt22}), i.e., $\delta_k=0$.
Therefore, $\delta_k=0$ for $\forall k \in \mathcal{K}$.
\end{IEEEproof}

With the aid of Lemma \ref{lem:lkn_delta2}, it is clear that the constraint (\ref{e:stRk}) is inactive and Equ.(\ref{e:gen_lkn_delta}) can be rewritten as follow.

%

\begin{equation}\label{e:lkn_delta2_simp}
  \Delta_{k,n} - \zeta_{k,n} = \frac{3 \varepsilon_k C_k^3 (R_k - \sum_{n=1}^{N} l_{k,n})^2}{T_k^2} - \frac{\Gamma C_k}{T_k - t_{tr}^*}, \forall k \in \mathcal{K}, \forall n \in \mathcal{N}
\end{equation}

For the specific $k$, the right-hand side of Equ.(\ref{e:lkn_delta2_simp}) is fixed, while its left-hand side is determined by the amount of offloading data size and channel conditions. To facilitate the analysis, we consider a special critical $\Delta_{k,n^\dag}$, which can make $\zeta_{k,n^\dag}$ exactly equal to $0$ while $n^\dag \notin \mathcal{A}_k$.
In this case, there is $\Delta_{k,n^\dag} = \frac{1}{h_{k,n^\dag}}$, and the corresponding channel gain $h_{k,n^\dag}$ called \textit{offloading channel gain condition}.

Next, we derive some properties about $\Delta_{k,n}$, $\forall k \in \mathcal{K}, \forall n \in \mathcal{N}$, and lemmas can be expected as follows.

\begin{lem} \label{lem:Delta_equal}
For $\forall i, j \in \mathcal{A}_k$, $\Delta_{k,i} = \Delta_{k,j}$.
\end{lem}
\begin{IEEEproof}
For $\forall n \in \mathcal{A}_k$, $\zeta_{k,n}=0$ according to constraint (\ref{e:lkn_kkt33}), and Equ.(\ref{e:lkn_delta2_simp}) can be simplified as follow
\begin{equation}\label{e:lkn_delta2_2}
  \Delta_{k,n} = \frac{3 \varepsilon_k C_k^3 (R_k - \sum_{n=1}^{N} l_{k,n})^2}{T_k^2} - \frac{\Gamma C_k}{T_k - t_{tr}^*},
\end{equation}
whose right-hand side is fixed for a specific $k$. For $i, j \in \mathcal{A}_k$, both $i$ and $j$ satisfy Equ.(\ref{e:lkn_delta2_2}), which means $\Delta_{k,i} = \Delta_{k,j}$. In addition, it is clear that $\Delta_{k,i} = \Delta_{k,j} = \Delta_{k,n^\dag}$.

Generally, $\{h_{k,n}\}$ is independent and different channel gains \cite{robust2016fan, dynamic2009Acharya}.
Therefore, it is almost certain that $\frac{l_{k,n}}{\tau_{k,n}^*}$ in $\Delta_{k,n}$ is different, which means the offloaded data size per bandwidth depends on channel conditions.
\end{IEEEproof}

\begin{lem} \label{lem:Delta_equal2}
For $\forall i \in \mathcal{A}_k$, $\forall j \notin \mathcal{A}_k$, $\Delta_{k,i} h_{k,j} \leq 1$.
\end{lem}
\begin{IEEEproof}
For $\forall i \in \mathcal{A}_k$, $\forall j \notin \mathcal{A}_k$, it can be seen that $l_{k,i} > 0$ and $l_{k,j} = 0$ according to definition of active channel set $\mathcal{A}_k$.
Therefore, Equ.(\ref{e:lkn_delta2_simp}) can be rewritten as follows for $i$ and $j$
\begin{align}
  \Delta_{k,i} &= \frac{3 \varepsilon_k C_k^3 (R_k - \sum_{n=1}^{N} l_{k,n})^2}{T_k^2} - \frac{\Gamma C_k}{T_k - t_{tr}^*}, \label{e:Delta_ij1}\\
  \Delta_{k,j} - \zeta_{k,j} &= \frac{3 \varepsilon_k C_k^3 (R_k - \sum_{n=1}^{N} l_{k,n})^2}{T_k^2} - \frac{\Gamma C_k}{T_k - t_{tr}^*}. \label{e:Delta_ij2}
\end{align}
The right-hand side of Equ.(\ref{e:Delta_ij1}) and Equ.(\ref{e:Delta_ij2}) is same for a specific $k$, thus $\Delta_{k,i}$ and $\Delta_{k,j}- \zeta_{k,j}$ are equal. From definition of $\Delta_{k,n}$, we can deduce that $\Delta_{k,i} = \frac{1}{h_{k,j}} - \zeta_{k,j}$, which is obviously that $\Delta_{k,i} h_{k,j} + \zeta_{k,j} h_{k,j} = 1$, thus $\Delta_{k,i} h_{k,j} \leq 1$.
\end{IEEEproof}

Lemma \ref{lem:Delta_equal} indicates that, for active channels, i.e, $n \in \mathcal{A}_k$, the offloading data size can be expressed as $l_{k,n} =  \tau_{k,n} t_{tr^*} ln(\Delta_{k,n} h_{k,n})$, and $\Delta_{k,n}$ are equal for $n \in \mathcal{A}_k$, which can be uniformly written as $\Delta_{k,n^\dag}$.
From Lemma \ref{lem:Delta_equal2}, it is clear that, for inactive channels, i.e, $n \notin \mathcal{A}_k$, the offloading data size is $0$ and $\Delta_{k,n^\dag} h_{k,n} \leq 1$.
For brevity, we use $\Delta_k$ to represent $\Delta_{k,n^\dag}$. Then, the offloading data size of $k$th MU can be simplified to
\begin{equation}\label{e:gen_lkn}
  l_{k,n} =  \tau_{k,n} t_{tr^*} ln\left[\max(\Delta_k h_{k,n}, 1)\right], \forall k \in \mathcal{K}, \forall n \in \mathcal{N}
\end{equation}
from which, we can know that $l_{k,n}$ is monotonically increasing with respect to $\Delta_k$.

As a Lagrange multiplier, $\Gamma$ should be equal to or larger than zero. Then, there are two possible cases: 1) $\Gamma = 0$; 2) $\Gamma > 0$. With mathematical transformations, for $k \in \mathcal{K}$ and $\forall n \in \mathcal{N}$, $\Gamma$ can be derived from Equ.(\ref{e:lkn_delta2_simp}) as $\Gamma = ( \frac{3 \varepsilon_k C_k^3 (R_k - \sum_{n=1}^{N} l_{k,n})^2}{T_k^2} + \zeta_{k,n} - \Delta_{k,n}) \frac{T_k - t_{tr}^*}{C_k}$.

1) When $\Gamma=0$, we can deduce that $\frac{3 \varepsilon_k C_k^3 (R_k - \sum_{n=1}^{N} l_{k,n})^2}{T_k^2} + \zeta_{k,n} - \Delta_{k,n} = 0$, which holds for $\forall n \in \mathcal{N}$ with a specific $k$.
For $n \in \mathcal{A}_k$, we can obtain $\frac{3 \varepsilon_k C_k^3 (R_k - \sum_{n=1}^{N} l_{k,n})^2}{T_k^2} = \Delta_k$. Substituting Equ.(\ref{e:gen_lkn}) into the above equation, $\Delta_k$ can be solved with bisection search.

2) When $\Gamma > 0$, we know that the constraint (\ref{e:Elkn_Fc}) is active from (\ref{e:lkn_kkt44}), namely $\sum_{k=1}^{K} \frac{C_k \sum_{n=1}^{N} l_{k,n}}{T_k - t_{tr}} - F_c =0$.
Substituting Equ.(\ref{e:gen_lkn}) into $\sum_{k=1}^{K} \frac{C_k \sum_{n=1}^{N} l_{k,n}}{T_k - t_{tr}^*} - F_c =0$, $\{\Delta_k\}$ can be found by linear programming.

With the obtained $\{\Delta_k\}$, the optimal offloading data size $\{l_{k,n}\}$ of above two cases can be solved by following Equ.(\ref{e:gen_lkn}) for $\forall k \in \mathcal{K}, \forall n \in \mathcal{N}$.
Finally, by comparing the two minimal achieved utilities for the case $\Gamma = 0$ and $\Gamma > 0$, the smaller achieved utility is the optimal utility of Problem \ref{p:problem_flk}, and the associated $\{l_{k,n}\}$ for $\forall k \in \mathcal{K}, \forall n \in \mathcal{N}$ is the optimal offloading data size.

Until now, for a given $t_{tr}$, we can find the optimal solution of Problem \ref{p:problem_flk} by solving Problem \ref{p:problem_tau} and Problem \ref{p:problem_lknfck} iteratively with a iterative algorithm, whose solving procedure can be summarized as follows.
\begin{algorithm}[H]
\caption{Iterative Algorithm to Solve Problem \ref{p:problem_flk} with given $t_{tr}$}
\label{a:pro_t}
\begin{algorithmic}[1]
\STATE Initialize $t_{tr}$, $f_{c_k}^{(0)}$, $l_{k,n}^{(0)}$. Let $\epsilon_1>0$ and $r=0$, and simplify Problem \ref{p:problem_flk}.
\REPEAT [Block coordinate descent method]
 \STATE Solve Problem \ref{p:problem_tau} for given $\{f_{c_k}^{(r)}, l_{k,n}^{(r)}\}$, and obtain the optimal $\{\tau_{k,n}^{(r+1)}\}$ from (\ref{e:opt_tau}), whose $\lambda_k$ is solved via bisection search.
 \STATE  Solve Problem \ref{p:problem_lknfck} for given $\{\tau_{k,n}^{r+1}\}$, obtain the optimal $\{l_{k,n}^{r+1}\}$ by comparing the two minimal achieved utilities for $\Gamma>0$ and $\Gamma=0$, and calculate the optimal $\{f_{c_k}^{r+1}\}$.
 \STATE Update $r=r+1$.
\UNTIL{Converge to the prescribed accuracy $\epsilon_1$.}


\end{algorithmic}
\end{algorithm}

Since Problem \ref{p:problem_flk} is jointly convex with respect to computation capacity $\{f_{c_k}\}$, offloading data size $\{l_{k,n}\}$ and subchannel bandwidth $\{\tau_{k,n}\}$ when $t_{tr}$ is given, solving Problem \ref{p:problem_tau} and Problem \ref{p:problem_lknfck} iteratively can guarantee the convergence to the optimal solution of Problem \ref{p:problem_flk} with given $t_{tr}$.

\subsection{Summary of the Optimal Solution for Problem \ref{p:problem}} \label{s:general_summary}
Based on the analysis in the preceding subsections, we propose Algorithm \ref{a:pro_general} to solved Problem \ref{p:problem} for solving the solution of $\{t_{tr},\bm{ f_l, f_c, l, \tau }\}$, where $t_{tr}^{\rm cur}$ and $E^*$ denote the current value of $t_{tr}$ and optimal value of the problem, respectively.

\begin{algorithm}[H]
\caption{Multi-user Multi-channel Offloading Strategy to Solve Problem \ref{p:problem} (MMOS)}
\label{a:pro_general}
\begin{algorithmic}[1]
\STATE Transform Problem \ref{p:problem} to Problem \ref{p:problem_flk}, set a step-size $\mu$ and initialize $t_{tr}^{\rm cur}=0$ and $E^*=E(t_{tr}^{\rm cur})$.
\WHILE {$t_{tr}^{\rm cur} < T_k$}
    \STATE Given current $t_{tr}^{\rm cur}$, solve Problem \ref{p:problem_flk} as in Algorithm \ref{a:pro_t} and obtain the current optimal solution of $E(t_{tr}^{\rm cur})$.
    \IF {$E(t_{tr}^{\rm cur}) < E^*$}
     \STATE Set $t_{tr}^{\rm *} = t_{tr}^{\rm cur}$,
     \STATE Set $E^* = E(t_{tr}^{\rm cur})$,
    \ENDIF
    \STATE Update $t_{tr}^{\rm cur} = t_{tr}^{\rm cur} + \mu$.
\ENDWHILE
\STATE Output $t_{tr}^*= \mathop{\arg\min}\limits_{t_{tr} \in[0, T_k)} E(t_{tr})$, the corresponding $\{f_{l_k}^*, f_{c_k}^*, l_{k,n}^*, \tau_{k,n}^* \}$ and $E(t_{tr}^*)$ are the optimal solution and  optimal utility of Problem \ref{p:problem}, respectively.
\end{algorithmic}
\end{algorithm}
Since Algorithm \ref{a:pro_t} can converge, the convergence of Algorithm \ref{a:pro_general} can be guaranteed by proving the convergence of loop in one-dimensional search.
Since the iteration times $r$ of one-dimensional search satisfies $\mu r \leq T_k$, i.e., $r \leq R_k / \mu$, thus the one-dimensional search will be completed within a finite number of times.
Thus the optimal solution of Problem \ref{p:problem_flk} can be found by exhaustive search, and the one-dimensional search in Algorithm \ref{a:pro_general} is convergent.

For Algorithm \ref{a:pro_general}, the complexity is composed of one-dimensional search of $t_{tr}$ and BCD method in Algorithm \ref{a:pro_t} for solving $\{\tau_{k,n}\}$, $\{l_{k,n}\}$ and $\{f_{c_k}\}$ in each iteration in while.

In one-dimensional search of $t_{tr}$, finding the optimal $t_{tr}^*$ has $R_k / \mu$ iterations at most, whose complexity is $O(R_k / \mu)$.
In Algorithm \ref{a:pro_t}, the complexity mainly comes from the bisection method for obtaining $\{\lambda_n\}$, $\{\tau_{k,n}\}$, the linear programming of $\{\Delta_k\}$, $\{l_{k,n}\}$, and the closed-form solution of $\{f_{c_k}\}$, which can be expressed as $O(N \rm{log}_2(1/\epsilon_2) + K N)$, $O(K^{3.5}+K N)$ and $O(K)$, respectively, where $\epsilon_2$ denotes the prescribed accuracy of bisection search.
Considering that the complexity of BCD iteration is $O(\rm{log}_2(1/\epsilon_1))$, the total computation complexity for Algorithm \ref{a:pro_t} is $O((N \rm{log}_2(1/\epsilon_2) + (2N+1)K + K^{3.5})\rm{log}_2(1/\epsilon_1))$.

Hence, the complexity of Algorithm \ref{a:pro_general} can be represented as $O((N \rm{log}_2(1/\epsilon) + (2N+1)K + K^{3.5})\rm{log}_2(1/\epsilon) (R_k / \mu))$
%

\section{Low-Complexity Solutions in Special Cases} \label{s:main_special}
To gain further insight of the structure of the optimal task offloading and resource allocation strategy, two representative special cases are worthy of consideration:
1) Multiple users share single channel;
2) Single user occupies all channels.
For the first case, though the solution approach is close to that of general case, there will lead to useful insights into optimal strategy.
For the second case, the optimal solution can be achieved in a low-complexity way and does not need to iterate repeatedly as in Algorithm \ref{a:pro_general}.

In the following subsections, we first study the multiple users share single channel and then consider the single user occupies all channel.

\subsection{When N=1, Multi-user Single Channel} \label{s:special_n1}
For the case of multiple users share single channel, multiple MUs compete for data offloading and resource allocation on a single channel, Problem \ref{p:problem} is simplified from multiple channels to single channel. 

For simplicity, variables can be simplified as shown: $l_k = \sum_{n=1}^{N} l_{k,n}$ and $\tau_k = \tau_{k,n}$ for $\forall k \in \mathcal{K}$.
Thus, the optimization variable of the Problem \ref{p:problem} is simplified to $\{t_{tr}, \{f_{l_k}\}, \{f_{c_k}\}, \{l_k\}, \{\tau_k\}\}$.
Since the property of ${f_{l_k}}$ in Problem \ref{p:problem} is not affected, similar analysis in Section \ref{s:general_local} can be applied, and the optimal local computational speed $f_{l_k}^*$ for $\forall k \in \mathcal{K}$ can be expressed as
$f_{l_k}^* = \frac{C_k (R_k - l_k)}{T_k}.$
With the optimal $f_{l_k}^*$, the problem can be equivalent to the problem below:

\begin{prob} \label{p:special_n1}
\begin{subequations}
\begin{align}
  \mathop{\min} \limits_{t_{tr}, \bm{ f_c, l, \tau}} \quad & \sum_{k=1}^{K} \omega_k E_k(t_{tr}, f_{c_k}, l_k, \tau_k)  \nonumber\\
  \text{s.t.}  \quad & \frac{C_k l_k}{f_{c_k}} \leq T_k - t_{tr}, \forall k \in \mathcal{K}, \label{e:spec_nT}\\
               \quad & \sum_{k=1}^{K} f_{c_k} \leq F_c, \label{e:spec_nFc}\\
               \quad & f_{c_k} \geq 0, \forall k \in \mathcal{K}, \label{e:spec_nfc}\\
               \quad & 0 \leq l_k \leq R_k, \forall k \in \mathcal{K}, \label{e:spec_nRk}\\
               \quad & \sum_{k=1}^{K} \tau_k \leq \tau, \label{e:spec_ntau}\\
               \quad & \tau_k \geq 0, \forall k \in \mathcal{K}, \label{e:spec_ntau2}
\end{align}
\end{subequations}
\end{prob}
where $E_k(t_{tr}, f_{c_k}, l_k, \tau_k)$ is expressed as
$E_k(t_{tr}, f_{c_k}, l_k, \tau_k) = \frac{\varepsilon_k C_k^3 (R_k - l_k)^3 }{T_k^2} + (e^{\frac{l_k}{\tau_k t_{tr}}} -1)\frac{\tau_k t_{tr}}{h_k}$.

Observing that Problem \ref{p:special_n1} has similar form with Problem \ref{p:problem_flk}, a similar procedure can be adopted for solving Problem \ref{p:special_n1}, i.e., the optimal transmission time $t_{tr}^*$ can be found by one-dimensional search and Problem \ref{p:special_n1} is a convex problem over $\{f_{c_k}, l_k, \tau_k\}$ with the given $t_{tr}$.
Thus, Problem \ref{p:special_n1} is block multi-convex for $\{f_{c_k}, l_k, \tau_k\}$ and can also be solved by applying BCD method with alternatively solving the following two subproblems, which are similar to Problem \ref{p:problem_tau} and Problem \ref{p:problem_lknfck}:

\begin{prob} \label{p:problem_spectau}
\begin{subequations}
\begin{align}
  \mathop{\min} \limits_{\bm{ \tau }} \quad &  \sum_{k=1}^{K} \omega_k E_k( \overline{f_{c_k}}, \overline{l_k}, \tau_k)  \nonumber\\
  \text{s.t.}  \quad & \text{Constraint} ~(\ref{e:spec_ntau}) , ~(\ref{e:spec_ntau2}). \nonumber
\end{align}
\end{subequations}
\end{prob}

\begin{prob} \label{p:problem_speclknfck}
\begin{subequations}
\begin{align}
  \mathop{\min} \limits_{\bm{ f_c, l }} \quad &  \sum_{k=1}^{K} \omega_k E_k( f_{c_k}, l_k, \overline{\tau_k})  \nonumber\\
  \text{s.t.}  \quad & \text{Constraints}~(\ref{e:spec_nT}) \text{--} (\ref{e:spec_nRk}). \nonumber
\end{align}
\end{subequations}
\end{prob}
Since there is only one channel, the constraints of multiple subchannels have become single channel constraints.

The Problem \ref{p:problem_spectau} has similar form to Problem \ref{p:problem_tau}, which is a convex problem satisfying Slater's condition. To analyze this problem, the KKT condition can be listed as follows:
\begin{subequations}
\begin{align}
   -\frac{t_{tr}^*}{h_k} \left[ e^{\frac{l_k}{\tau_k t_{tr}^*}} (\frac{l_k}{\tau_k t_{tr}^*} -1) +1 \right] + \lambda^* - v_k^* &= 0, \forall k \in \mathcal{K} \label{e:spec_tau_kkt1}\\
   \lambda^* (\sum_{k=1}^{K} \tau_k - \tau) &= 0, \label{e:spec_tau_kkt2}\\
   v_k^* \tau_k &= 0, \forall k \in \mathcal{K} \label{e:spec_tau_kkt3}\\
   \text{Constraint} ~(\ref{e:spec_ntau}) ,& ~(\ref{e:spec_ntau2}). \nonumber
\end{align}
\end{subequations}
in which $\lambda^*$ and $v_k^*$ are non-negative Lagrange multipliers associated with constraints (\ref{e:spec_ntau}) and (\ref{e:spec_ntau2}), respectively.

Similar to Section \ref{s:general_channel}, the optimal channel bandwidth allocation $\forall k \in \mathcal{K}$ can be derived as
\begin{equation}\label{e:spec_opt_tau}
  \tau_k^* =
  \begin{cases}
    \frac{l_k}{t_{tr}^* \left( W_k + 1 \right)}, & W_k > -1 \\
    0, & \mbox{otherwise}.
  \end{cases}
\end{equation}
where $W_k = W_0 \left(- \frac{1}{e} [\frac{h_k}{t_{tr}^*} (v_k^* - \lambda^*) +1]\right)$, and $W_0(\cdot)$ denotes the $0$th branch of the Lambert W function.
Note that, when $W_k = -1$, there is $l_k = 0$, and the corresponding $\tau_k$ should be $0$.

To obtain the optimal channel bandwidth $\tau_k^*$ via Equ.(\ref{e:spec_opt_tau}) requires computing the Lagrange multipliers $\lambda^*$ and $v_k^*$ by solving constraint (\ref{e:spec_tau_kkt2}) and constraint (\ref{e:spec_tau_kkt3}).
Define the active user set $\mathcal{A} \triangleq \{k|l_k >0, \tau_k >0, k \in \mathcal{K}\}$, which means the user who offloads.
For the KKT condition, the following lemmas are in order.
\begin{lem} \label{lem:spec_lkn_lambda}
$\lambda^*>0$.
\end{lem}
\begin{IEEEproof}
See the proof of Lemma \ref{lem:gen_lkn_lambda}.
%
%
\end{IEEEproof}

With the aid of Lemma \ref{lem:spec_lkn_lambda}, the Lagrange multiplier $\lambda^*$ and optimal channel bandwidth $\{\tau_k^*\}$ can be solved with the same solution approach developed in Section \ref{s:general_channel}, and details are omitted here for brevity.


Next, we solve the optimal computation capacity allocation $f_{c_k}$ and offloading data size $l_k$.
For Problem \ref{p:problem_speclknfck}, the optimal allocation of computation capacity of edge server $f_{c_k}^*$ can be derived from constraint (\ref{e:spec_nT}), which is similar to the derivation in Section \ref{s:general_offloading}. Therefore, the closed-form solution of optimal allocation of computation capacity of edge server is $f_{c_k}^*=\frac{C_k l_k}{T_k - t_{tr}^*}$.


With optimal $f_{c_k}^*$, Problem \ref{p:problem_speclknfck} is a convex problem with respect to $l_k$, and satisfies Slater's condition. To analyze this problem, the KKT condition can be listed as follows:
\begin{subequations}
\begin{align}
  -\frac{3 \varepsilon_k C_k^3 (R_k - l_k)^2}{T_k^2} + \frac{1}{h_k} e^{\frac{l_k}{\tau_k^* t_{tr}^*}} + \frac{\Gamma^* C_k}{T_k - t_{tr}^*} + \delta_k^* - \zeta_k^* &= 0, \forall k \in \mathcal{K} \label{e:spec_lkn_kkt1}\\
  \Gamma^* (\sum_{k=1}^{K} \frac{C_k l_k}{T_k - t_{tr}^*} - F_c) &= 0 \label{e:spec_lkn_kkt2} \\
  \delta_k^* (l_k - R_k) &= 0, \forall k \in \mathcal{K} \label{e:spec_lkn_kkt3}\\
  \zeta_k^* l_k &= 0, \forall k \in \mathcal{K} \label{e:spec_lkn_kkt4}\\
  \text{Constraints}~(\ref{e:spec_nT}) ,& ~(\ref{e:spec_nRk}).   \nonumber
\end{align}
\end{subequations}
where $\Gamma^*$, $\delta_k^*$ and $\zeta_k^*$ are non-negative Lagrange multipliers associated with constraints $\sum_{k=1}^{K} \frac{C_k l_k}{T_k - t_{tr}^*} \leq F_c$, $l_k \leq R_k$ and $l_k \geq 0$, respectively.

By analyzing the KKT condition, a lemma can be expected as follow.
\begin{lem} \label{lem:spec_lkn_delta}
For $k \in \mathcal{K}$, $\delta_k^*=0$.
\end{lem}
\begin{IEEEproof}
For $\forall k \in \mathcal{A}$, $\zeta_k^*=0$ according to constraint (\ref{e:spec_lkn_kkt4}), and constraint (\ref{e:spec_lkn_kkt1}) can be rewritten as
\begin{equation}\label{e:spec_lk_delta}
  \frac{3 \varepsilon_k C_k^3 (R_k - l_k)^2}{T_k^2} = \frac{1}{h_k} e^{\frac{l_k}{\tau_k^* t_{tr}^*}} + \frac{\Gamma^* C_k}{T_k - t_{tr}^*} + \delta_k^*, \forall k \in \mathcal{K}
\end{equation}
As Lagrange multipliers, $\Gamma^*$ and $\delta_k^*$ should be equal to or larger than $0$. Observed that when Equ.(\ref{e:spec_lk_delta}) holds, its right-hand side is larger than $0$, and the left-hand side should also be larger than $0$, namely $l_k<R_k$.
For $l_k<R_k$, we can derive that $\delta_k^*=0$ from constraint (\ref{e:spec_lkn_kkt3}).
Therefore, $\delta_k^*=0$ for $k \in \mathcal{K}$, and $l_k \in [0, R_k)$, which means constraint $l_k \leq R_k$ is inactive, thus, MUs will not offload all tasks.
%
\end{IEEEproof}

With the aid of Lemma \ref{lem:spec_lkn_delta}, it is clear that MUs complete all or part of computation tasks locally. Then constraint (\ref{e:spec_lkn_kkt1}) can be rewritten as:
\begin{equation}\label{e:spec_lkn_nodelta}
  \frac{\Gamma^* C_k}{T_k - t_{tr}^*} = \frac{3 \varepsilon_k C_k^3 (R_k - l_k)^2}{T_k^2} - \frac{1}{h_k} e^{\frac{l_k}{\tau_k^* t_{tr}^*}} + \zeta_k^*, \forall k \in \mathcal{K}
\end{equation}
from which, we can deduce that $\Gamma^*$ is monotonically decreasing with respect to $l_k$.

For the Lagrange multiplier $\Gamma^*$, there are two possible cases: 1) $\Gamma^*=0$, and 2) $\Gamma^*>0$. Besides, $\Gamma^*$ can be expressed as $\Gamma^* = \left[ \frac{3 \varepsilon_k C_k^3 (R_k - l_k)^2}{T_k^2} - \frac{1}{h_k} e^{\frac{l_k}{\tau_k^* t_{tr}^*}} + \zeta_k^* \right] \frac{T_k - t_{tr}^*}{C_k}$ from Equ.(\ref{e:spec_lkn_nodelta}), for $\forall k \in \mathcal{K}$.

1) When $\Gamma^*=0$, we can know that $\frac{3 \varepsilon_k C_k^3 (R_k - l_k)^2}{T_k^2} - \frac{1}{h_k} e^{\frac{l_k}{\tau_k^* t_{tr}^*}} + \zeta_k^* = 0$, for $k \in \mathcal{K}$.
For $k \notin \mathcal{A}$, thus, $\frac{3 \varepsilon_k C_k^3 R_k^2}{T_k^2} < \frac{1}{h_k}$, the optimal offloading data size $\{l_k\}$ is $0$.
For $k \in \mathcal{A}$, we can simplify the equation to $\frac{3 \varepsilon_k C_k^3 (R_k - l_k)^2}{T_k^2} = \frac{1}{h_k} e^{\frac{l_k}{\tau_k^* t_{tr}^*}}$ and the optimal offloading data size $\{l_k\}$ can be solved by bisection search.

2) When $\Gamma^*>0$, the constraint (\ref{e:spec_nT}) is active according to constraint (\ref{e:spec_lkn_kkt2}), i.e., $\sum_{k=1}^{K} \frac{C_k l_k}{T_k - t_{tr}^*} = F_c$. Then the optimal offloading data size $\{l_k\}$ can be solved by linear programming.

Comparing the two minimal achieved utilities of above two cases, the smaller achieved utility is selected as the optimal utility of origin Problem, and the associated offloading data size $\{l_k\}$ for $k \in \mathcal{A}$ that can achieve the optimal utility are the optimal solution.


Similar with the discussion in Section \ref{s:general_summary}, the algorithm for solving this problem is also convergent, and the computation complexity for this case is the same as the computation complexity of Algorithm \ref{a:pro_general}.

\subsection{When K=1, Single user Multi-Channel} \label{s:special_k1}
For the case of single user occupies all channels, the only MU occupies all computation capacity and channel bandwidth, and Problem \ref{p:problem} can be simplified to the issue of data offloading and transmission time allocation. 

In this case, variables can be simplified as $f_{l_k} = f_l$, $f_{c_k} = F_c$, $\tau_{k,n} = \tau$, for $\forall n \in N$, namely, the optimization variable of the Problem \ref{p:problem} is simplified to $\{t_{tr}, f_l, \{l_{k,n}\}\}$.
Since the nature of $f_l$ in Problem \ref{p:problem} is not affected, similar analysis in Section \ref{s:general_local} can be applied, and Problem \ref{p:problem_flk} can be equivalent to the problem below:
\begin{prob} \label{p:special_k1}
\begin{subequations}
\begin{align}
  \mathop{\min} \limits_{t_{tr}, \{l_{k,n}\}} \quad & E(t_{tr},l_{k,n})   \nonumber\\
  \text{s.t.}  \quad & \frac{C_k \sum_{n=1}^{N}l_{k,n}}{F_c} \leq T_k - t_{tr}, \forall k \in \mathcal{K}, \label{e:t_tr}\\
               \quad & \sum_{n=1}^{N}l_{k,n} \leq R_k, \forall k \in \mathcal{K},\\
               \quad & l_{k,n} \geq 0, \forall k \in \mathcal{K}, \forall n \in \mathcal{N},
\end{align}
\end{subequations}
\end{prob}
where $E(t_{tr},l_{k,n})$ is expressed as
$E(t_{tr},l_{k,n}) = \frac{\varepsilon_k C_k^3 (R_k - \sum_{n=1}^{N}l_{k,n})^3 }{T_k^2} + \sum_{n=1}^{N} (e^{\frac{l_{k,n}}{\tau t_{tr}}} -1)\frac{\tau t_{tr}}{h_{k,n}}$.

Next, we will analyze the special properties of Problem \ref{p:special_k1}, and study a low-complexity algorithm with a closed-form optimal solution for the problem.
For Problem \ref{p:special_k1}, the optimal $t_{tr}$ can be given in the following lemma.
\begin{lem} \label{lem:tr}
With given $l_{k,n}$, the optimal transmission time $t_{tr}^*$ of Problem \ref{p:special_k1} can be deduced as follow.
\begin{equation}\label{e:opt_t}
  t_{tr}^*(l_{k,n}) = T_k - \frac{C_k \sum_{n=1}^{N} l_{k,n}}{F_c},
\end{equation}
\end{lem}
\begin{IEEEproof}
Define $f(t_{tr},l_{k,n}) \triangleq (e^{\frac{l_{k,n}}{\tau t_{tr}}} -1)\frac{\tau t_{tr}}{h_{k,n}}$, we investigate the first-order and second-order partial derivatives of $f(t_{tr},l_{k,n})$ with $t_{tr}$, there are $\frac{\partial {f(t_{tr},l_{k,n})}}{\partial{t_{tr}}} = \frac{\tau}{h_{k,n}} [e^{\frac{l_{k,n}}{\tau t_{tr}}}(1-\frac{l_{k,n}}{\tau t_{tr}}) -1]$, and $\frac{\partial^2 f(t_{tr},l_{k,n})}{\partial t_{tr}^2} =  \frac{l_{k,n}^2 e^{\frac{l_{k,n}}{\tau t_{tr}}}}{h_{k,n} \tau t_{tr}^3}$. When $l_{k,n} \geq 0$ and $t_{tr} \geq 0$, there always exists $\frac{\partial^2 f(t_{tr},l_{k,n})}{\partial t_{tr}^2} \geq 0$, which indicates $\frac{\partial {f(t_{tr},l_{k,n})}}{\partial{t_{tr}}}$ keep on increasing with $t_{tr}$.
In addition, $\lim \limits_{t_{tr} \to 0^+} \frac{\partial {f(t_{tr},l_{k,n})}}{\partial{t_{tr}}} = -\infty$ and $\lim \limits_{t_{tr} \to \infty} \frac{\partial {f(t_{tr},l_{k,n})}}{\partial{t_{tr}}} = 0$. Hence, $\frac{\partial {f(t_{tr},l_{k,n})}}{\partial{t_{tr}}}$ is always less than $0$ for $t_{tr} \geq 0$, which proves the decreasing monotonicity of $f(t_{tr},l_{k,n})$ with $t_{tr}$.

To achieve the minimum value of $f(t_{tr},l_{k,n})$, $t_{tr}$ should be as large as possible, in which case the constraints (\ref{e:t_tr}) become active, thus $t_{tr}^*(l_{k,n}) = T_k - \frac{C_k \sum_{n=1}^{N} l_{k,n}}{F_c}$.
Considering that the objective function of Problem \ref{p:special_k1} and $f(t_{tr},l_{k,n})$ have same monotonicity with respect to $t_{tr}$, the minimal utility of Problem \ref{p:special_k1} is also achieved at $t_{tr}^*(l_{k,n})$.
Note that $t_{tr}^* \geq 0$ means $\sum_{n=1}^{N}l_{k,n} \leq \frac{T_k F_c}{C_k}$.
\end{IEEEproof}

Substituting $t_{tr}^*$ in Equ.(\ref{e:opt_t}) into the objective function of Problem \ref{p:special_k1}, and we can obtain
$E(l_{k,n}) =  \frac{\varepsilon_k C_k^3 (R_k - \sum_{n=1}^{N}l_{k,n})^3 }{T_k^2} + 
  \sum_{n=1}^{N} (e^{\frac{l_{k,n}}{\tau (T_k - \frac{C_k \sum_{n=1}^{N} l_{k,n}}{F_c})}} -1)\frac{\tau (T_k - \frac{C_k \sum_{n=1}^{N} l_{k,n}}{F_c})}{h_{k,n}}$.
Define $l_{k,t} \triangleq \sum_{n=1}^{N}l_{k,n}$, via changing variables $\{l_{k,n},\sum_{n=1}^{N}l_{k,n}\}$ into $\{l_{k,n}, l_{k,t}\}$, $E(l_{k,n})$ can be converted to $E(l_{k,n},l_{k,t})$, which can be expressed as
\begin{equation}\label{e:Elkn_lkt}
\begin{array}{ll}
  E(l_{k,n},l_{k,t}) = 
  \frac{\varepsilon_k C_k^3 (R_k - l_{k,t})^3 }{T_k^2} 
  + \sum_{n=1}^{N} (e^{\frac{l_{k,n}}{\tau (T_k - \frac{C_k l_{k,t}}{F_c})}} -1)\frac{\tau (T_k - \frac{C_k l_{k,t}}{F_c})}{h_{k,n}}.
\end{array}
\end{equation}


Observed that $E(l_{k,n},l_{k,t})$ could be solved by \textit{bilevel optimization} \cite{Encyclopedia_Optimization,joint_bilevel}, in which the lower level problem is to optimize subchannel offloading $\{l_{k,n}\}$ with fixed total offloading $l_{k,t}$, while the upper level problem is to optimize total offloading $l_{k,t}$.
Specifically, the lower level problem is
\begin{prob} \label{p:special_lkn}
\begin{subequations}
\begin{align}
   F(l_{k,t}) \triangleq \mathop{\min} \limits_{\{l_{k,n}\}} \quad & E_{\rm{low}}(l_{k,n}, l_{k,t}) \nonumber\\
   \text{s.t.} \quad & \sum_{n=1}^{N} l_{k,n} \geq l_{k,t}, \forall k \in \mathcal{K}, \label{e:lkn_1}
\end{align}
\end{subequations}
\end{prob}
where $E_{\rm{low}}(l_{k,n}, l_{k,t})$ is expressed as $E_{\rm{low}}(l_{k,n}, l_{k,t}) = \sum_{n=1}^{N} (e^{\frac{l_{k,n}}{\tau (T_k - \frac{C_k l_{k,t}}{F_c})}} -1)\frac{\tau (T_k - \frac{C_k l_{k,t}}{F_c})}{h_{k,n}}$.

Then, the upper level problem is
\begin{prob} \label{p:special_lkt}
\begin{subequations}
\begin{align}
  \mathop{\min} \limits_{l_{k,t}} \quad &  E_{\rm{up}}(l_{k,t})   \nonumber\\
  \text{s.t.}  \quad & l_{k,t} \leq \frac{T_k F_c}{C_k}, \forall k \in \mathcal{K}, \\
               \quad & 0 \leq l_{k,t} \leq R_k, \forall k \in \mathcal{K},
\end{align}
\end{subequations}
\end{prob}
where $E_{\rm{up}}(l_{k,t})$ is expressed as $E_{\rm{up}}(l_{k,t}) = \frac{\varepsilon_k C_k^3 (R_k - l_{k,t})^3 }{T_k^2} + F(l_{k,t})$.

For the lower level problem, i.e., Problem \ref{p:special_lkn}, it can be easily proved that the feasible region is convex. Then, we derive the following lemma to show the convexity of Problem \ref{p:special_lkn}.
\begin{lem} \label{lem:spec2_jointconvex}
$E_{\rm{low}}(l_{k,n}, l_{k,t})$ is jointly convex function with respect to $l_{k,n}$ and $l_{k,t}$, $\forall n \in \mathcal{N}$.
\end{lem}
\begin{IEEEproof}
Since $E_{\rm{low}}(l_{k,n}, l_{k,t})$ is separable with subchannel $n$, it is equivalent to prove that the corresponding part of each subchannel in $E_{\rm{low}}(l_{k,n}, l_{k,t})$ is jointly convex with respect to $l_{k,n}$ and $l_{k,t}$.
For $\forall n \in \mathcal{N}$, the Hessian matrix $\mathcal{H}$ of $E_{\rm{low}}(l_{k,n}, l_{k,t})$ can be obtained as
\begin{equation}\label{e:spec2_hessian}
  \mathcal{H} = \frac{e^{\frac{l_{k,n}}{\tau \left(T_k - \frac{C_k l_{k,t}}{F_c}\right)}}}{\tau h_{k,n} \left(T_k - \frac{C_k l_{k,t}}{F_c}\right)^2}
  \left[ \begin{array}{cc}
           \left(T_k - \frac{C_k l_{k,t}}{F_c}\right) & \frac{C_k l_{k,n}}{F_c} \\
           \frac{C_k l_{k,n}}{F_c} & \frac{(C_k l_{k,n})^2}{F_c^2 \left(T_k - \frac{C_k l_{k,t}}{F_c}\right)}
         \end{array}
  \right]
\end{equation}
with eigenvalues $0$ and $e^{\frac{l_{k,n}}{\tau \left(T_k - \frac{C_k l_{k,t}}{F_c}\right)}} \frac{\left[ \left(T_k - \frac{C_k l_{k,t}}{F_c}\right)^2 + \left(\frac{C_k l_{k,n}}{F_c}\right)^2 \right]}{\tau h_{k,n} \left(T_k - \frac{C_k l_{k,t}}{F_c}\right)^3}$. It can be derived that all the eigenvalues are not less than $0$, thus the Hessian matrix $\mathcal{H}$ is a semipositive definite matrix. Hence, $E_{\rm{low}}(l_{k,n}, l_{k,t})$ is jointly convex function with respect to $l_{k,n}$ and $l_{k,t}$, $\forall n \in \mathcal{N}$.
\end{IEEEproof}


\begin{lem} \label{lem:spec2_flkt}
Function $F(l_{k,t})$ is convex with respect to $l_{k,t}$.
\end{lem}
\begin{IEEEproof}
Suppose $\exists l_{k,t}^\dagger, l_{k,t}^\ddagger \in [0, R_k]$, and the optimal solution of Problem \ref{p:special_lkn} with $l_{k,t}=l_{k,t}^\dagger$ and $l_{k,t}=l_{k,t}^\ddagger$ are $\{l_{k,n}^\dagger\}$ and $\{l_{k,n}^\ddagger\}$, respectively.
Since $E_{\rm{low}}(l_{k,n}, l_{k,t})$ is jointly convex function with respect to $l_{k,n}$ and $l_{k,t}$, we have

\begin{equation} \label{e:spec2_F_deduce}
\begin{array}{cl}
\alpha F(l_{k,t}^\dagger) + (1-\alpha) F(l_{k,t}^\ddagger) &=  \inf \limits_{\{l_{k,n}\}} \alpha E_{\rm{low}}(l_{k,n}, l_{k,t}^{\dagger}) + \inf \limits_{\{l_{k,n}\}} (1-\alpha) E_{\rm{low}}(l_{k,n}, l_{k,t}^{\ddagger})\\
&= \alpha \sum_{n=1}^{N} (e^{\frac{l_{k,n}^\dagger}{\tau (T_k - \frac{C_k l_{k,t}^\dagger}{F_c})}} -1)\frac{\tau (T_k - \frac{C_k l_{k,t}^\dagger}{F_c})}{h_{k,n}} + \\
& ~~~~~~~~~~~~~~~~~~~~~(1-\alpha)  \sum_{n=1}^{N} (e^{\frac{l_{k,n}^\ddagger}{\tau (T_k - \frac{C_k l_{k,t}^\ddagger}{F_c})}} -1)\frac{\tau (T_k - \frac{C_k l_{k,t}^\ddagger}{F_c})}{h_{k,n}}\\
& \geq \sum_{n=1}^{N} (e^{\frac{\alpha l_{k,n}^\dagger + (1-\alpha) l_{k,n}^\ddagger}{\tau (T_k - \frac{C_k \left[\alpha l_{k,t}^{\dagger} + (1-\alpha) l_{k,t}^{\ddagger}\right]}{F_c})}} -1) \frac{\tau (T_k - \frac{C_k \left[\alpha l_{k,t}^{\dagger} + (1-\alpha) l_{k,t}^{\ddagger}\right]}{F_c})}{h_{k,n}}\\
& \geq \inf \limits_{\{l_{k,n}\}} \sum_{n=1}^{N} (e^{\frac{l_{k,n}}{\tau (T_k - \frac{C_k \left[\alpha l_{k,t}^{\dagger} + (1-\alpha) l_{k,t}^{\ddagger}\right]}{F_c})}} -1) \frac{\tau (T_k - \frac{C_k \left[\alpha l_{k,t}^{\dagger} + (1-\alpha) l_{k,t}^{\ddagger}\right]}{F_c})}{h_{k,n}}\\
& = F(\alpha l_{k,t}^{\dagger} + (1-\alpha) l_{k,t}^{\ddagger})
\end{array}
\end{equation}
where the first inequality can be derived from Lemma \ref{lem:spec2_jointconvex} and the last inequality comes from the fact that the optimal solution of Problem \ref{p:special_lkn} when $l_{k,t} = \alpha l_{k,t}^{* \dagger} + (1-\alpha) l_{k,t}^{* \ddagger}$ is the minimum.
Therefore, $F(l_{k,t})$ is convex with respect to $l_{k,t}$.
\end{IEEEproof}

With the aid of Lemma \ref{lem:spec2_flkt}, we know that Problem \ref{p:special_lkn} is a convex problem, which can be solved by existing methods. Instead of numerical optimization methods like subgradient method and interior point algorithm, we use KKT conditions to analysis the special properties of the Problem \ref{p:special_lkn}.
Thus, the optimal $\{l_{k,n}\}$ for given $\{l_{k,t}\}$ can be obtained.

Problem \ref{p:special_lkn} satisfies Slater's condition. Then, the KKT condition that serves as a sufficient and necessary condition for the optimal solution, which can be shown as follow \cite{Boyd}.
\begin{subequations}
\begin{eqnarray}
   \frac{1}{h_{k,n}} e^{\frac{l_{k,n}}{\tau (T_k - \frac{C_k l_{k,t}}{F_c})}} - \lambda_k^* = 0, \forall k \in \mathcal{K}, \forall n \in \mathcal{N} \label{e:speclkn_kkt1}\\
   \lambda_k^* (\sum_{n=1}^{N} l_{k,n} - l_{k,t}) = 0, \forall k \in \mathcal{K} \label{e:speclkn_kkt2}\\
   \text{Constraint}~(\ref{e:lkn_1})\nonumber
\end{eqnarray}
\end{subequations}
in which $\lambda_k^*$ is nonnegative Lagrange multipliers associated with constraints $\sum_{n=1}^{N} l_{k,n} \geq l_{k,t}$.
According to constraint (\ref{e:speclkn_kkt1}), it can be deduced that there is feasible solution when $\lambda_k^* \geq \frac{1}{h_{k,n}}$, and the optimal subchannel offloading strategy can be derived as follow
\begin{equation}\label{e:speclkn_optimal}
  l_{k,n}^* = \tau \left(T_k - \frac{C_k l_{k,t}}{F_c} \right) ln(\lambda_k^* h_{k,n})
\end{equation}
where $\lambda_k^* \geq \frac{1}{h_{k,n}}$.

Obtaining the optimal solution of subchannel offloading strategy, we can deduce that $\lambda_k^* > 0$ in constraint (\ref{e:speclkn_kkt1}), which means $\sum_{n=1}^{N} l_{k,n} - l_{k,t} = 0$ in constraint (\ref{e:speclkn_kkt2}).
Substitute $l_{k,n}^*$ in Equ.(\ref{e:speclkn_optimal}) into $\sum_{n=1}^{N} l_{k,n} - l_{k,t} = 0$, we can get an equation about $\lambda_k^*$, which can be derived as $\lambda_k^* = \left[\frac{1}{\prod_{n=1}^{N} h_{k,n}} e^{\frac{l_{k,t}}{\tau (T_k - \frac{C_k l_{k,t}}{F_c})}}\right]$
for $\forall k \in \mathcal{K}$.
With the solved $\lambda_k^*$, the optimal solution of $\{l_{k,n}\}$ for the given $l_{k,t}$ can be obtained.

By the obtained $\{l_{k,n}^*\}$, the objective function of upper level problem (i.e., Problem \ref{p:special_lkt}) can be converted to $ E_{\rm{up}}(l_{k,t}) = \frac{\varepsilon_k C_k^3 (R_k - l_{k,t})^3 }{T_k^2} + \tau (T_k - \frac{C_k l_{k,t}}{F_c}) \sum_{n=1}^{N} \left(\lambda_k^* - \frac{1}{h_{k,n}} \right) $, whose property can be deduced as follow.

\begin{lem} \label{lem:spec2_up}
$E_{\rm{up}}(l_{k,t})$ is a convex function.
\end{lem}
\begin{IEEEproof}
The second-order partial derivative is given by

$\frac{\partial^2 {E_{\rm{up}}(l_{k,t})}}{\partial{l_{k,t}^2}} = \frac{6\varepsilon_k C_k^3 (R_k - l_{k,t})}{T_k^2} + \frac{T_k^2}{N \tau (\prod_{n=1}^{N} h_{k,n})^{\frac{1}{N}} (T_k - \frac{C_k l_{k,t}}{F_c})^3} e^{\frac{l_{k,t}}{N \tau (T_k - \frac{C_k l_{k,t}}{F_c})}}$.
Since $T_k - \frac{C_k l_{k,t}}{F_c} > 0$ and $R_k - l_{k,t} >0$, there always exists $\frac{\partial^2 {E_{\rm{up}}(l_{k,t})}}{\partial{l_{k,t}^2}} \geq 0$, which means that $E_{\rm{up}}(l_{k,t})$ is a convex function.
\end{IEEEproof}


From Lemma \ref{lem:spec2_up}, it can be seen that the objective function of Problem \ref{p:special_lkt} is convex within the given interval, which means that the function is unimodal \cite{Boyd}, i.e., the function contains only one optimal point in the given bounded interval.
For the bounded $l_{k,t}$ in Problem \ref{p:special_lkt}, \textit{Golden Section Search Method} \cite{introductin2013Chong} can be adopted to find the optimal $l_{k,t}$, and Algorithm \ref{a:spec2_golden} details the procedure, where $\epsilon_g$ denotes the maximum tolerance.

\begin{algorithm}[H]
\caption{Golden Section Search.}
\label{a:spec2_golden}
\begin{algorithmic}[1]
\STATE Initialize $\epsilon_g >0$, $l_{k,t}^{\rm{low}}=0$ and $l_{k,t}^{\rm{up}} = \min\{\frac{T_k F_c}{C_k}, R_k\}$.
\STATE Determine two intermediate points $l_{k,t}^{\rm{mid1}}$ and $l_{k,t}^{\rm{mid2}}$, set $l_{k,t}^{\rm{mid1}} = l_{k,t}^{\rm{up}} - d$, and $l_{k,t}^{\rm{mid2}} = l_{k,t}^{\rm{low}} + d$,
where $d = \frac{\sqrt{5}-1}{2} ( l_{k,t}^{\rm{up}} -  l_{k,t}^{\rm{low}})$.
\REPEAT
    \STATE Obtain the current value of $E_{up}(l_{k,t}^{\rm{mid1}})$ and $E_{up}(l_{k,t}^{\rm{mid2}})$.
    \IF  {$E_{\rm{up}}(l_{k,t}^{\rm{mid1}}) > E_{\rm{up}}(l_{k,t}^{\rm{mid2}})$}
    \STATE Update $l_{k,t}^{\rm{low}} = l_{k,t}^{\rm{mid1}}$, $l_{k,t}^{\rm{mid1}} = l_{k,t}^{\rm{mid2}}$ and $l_{k,t}^{\rm{mid2}} = l_{k,t}^{\rm{low}} + \frac{\sqrt{5}-1}{2} ( l_{k,t}^{\rm{up}} -  l_{k,t}^{\rm{low}})$.
    \ELSE
    \STATE Update $l_{k,t}^{\rm{up}} = l_{k,t}^{\rm{mid2}}$, $l_{k,t}^{\rm{mid2}} = l_{k,t}^{\rm{mid1}}$ and $l_{k,t}^{\rm{mid1}} = l_{k,t}^{\rm{up}} - \frac{\sqrt{5}-1}{2} ( l_{k,t}^{\rm{up}} -  l_{k,t}^{\rm{low}})$.
    \ENDIF
    \UNTIL {$|l_{k,t}^{\rm{up}} -  l_{k,t}^{\rm{low}}| \leq \epsilon_g$}
\STATE  Output $\frac{l_{k,t}^{\rm{up}} +  l_{k,t}^{\rm{low}}}{2}$ as the optimal solution of $l_{k,t}$.
\end{algorithmic}
\end{algorithm}

By now, we have solve Problem \ref{p:special_k1}. The pseudo code of the method is expressed in Algorithm \ref{a:special_k1}.
\begin{algorithm}[H]
\caption{Single User Multi-channel Offloading Strategy to Solve Problem \ref{p:special_k1}.}
\label{a:special_k1}
\begin{algorithmic}[1]
\STATE Transform Problem\ref{p:problem_flk} to Problem \ref{p:special_k1}.
\STATE Obtain $t_{tr}^*$ from Lemma \ref{lem:tr}.
\STATE Changing variables into $\{l_{k,n},l_{k,t}\}$, transform Problem \ref{p:special_k1} into a bilevel problem with $t_{tr}^*$, where the upper level is Problem \ref{p:special_lkn} and the lower level is Problem \ref{p:special_lkt}.
\STATE Through analysis, we deduce $\lambda_k^*$ and obtain $\{l_{k,n}^*\}$ from (\ref{e:speclkn_optimal}) via Problem \ref{p:special_lkn}.
\STATE With obtained $\lambda^*_k$ and $\{l_{k,n}^*\}$, perform Algorithm \ref{a:spec2_golden} for solving optimal $\{l_{k,t}\}$ of Problem \ref{p:special_lkt}.
\end{algorithmic}
\end{algorithm}

The convergence of Algorithm \ref{a:special_k1} can be guaranteed by proving the convergence of golden section search. Since the iteration times $r$ of golden section search satisfies $0.618^r < R_k$, i.e., $r < \rm{log}_{0.618} R_k$, thus the golden section search will be completed after a finite number of times. Hence,  Algorithm \ref{a:special_k1} can converge to the optimal solution of Problem \ref{p:special_k1}.

For Algorithm \ref{a:special_k1}, the complexity is mainly comes from solving for $\{l_{k,n}\}$ and the golden section search of $l_{k,t}$. The complexity of semi-closed form solution is $O(K N)$, while the iteration complexity of golden section search is $O(K \rm{log}_{0.618} R_k)$, and each iteration has the complexity $O()$.
Therefore, the total computation complexity for Algorithm \ref{a:special_k1} is $O(K (N + \rm{log}_{0.618} R_k))$
%

%
%

\section{Conclusion} \label{s:main_conclusion}
In this paper, we have investigated the joint radio and computation resource management problem for a multi-user partial offloading MEC system.
The optimization problem, whose targets at minimizing the weighted sum of mobile energy consumption over multi-channel while limiting the computation capacity of both edge server and MUs, is formulated and transformed into a convex optimization problem equivalently.
We first consider the general case of multi-channel offloading resource management for multi-users under independent latency constraints, and deduce the optimal strategy by using the BCD method.
%
To obtain more insights into the structure of the optimal strategy, we further investigate two special cases.
For both cases, the energy consumption minimization problem is decomposed to subproblems, whose special properties are found and low complexity global optimal solutions are given.
%
This research should provide helpful insights for radio and computation resource management for multi-user MEC system under independent latency constraints.


\ifCLASSOPTIONcaptionsoff
  \newpage
\fi

\end{document}